\pdfoutput=1

\documentclass[11pt,a4paper]{article}

\usepackage{jheppub}
\usepackage{amsmath}
\usepackage{amssymb}
\usepackage{appendix}
\usepackage{multirow}
\usepackage{rotating}
\usepackage{diagbox}
\usepackage{float}
\usepackage{xcolor}
\usepackage{cancel}
\usepackage{tabularx}

\usepackage{color}

\def\stat{({\rm stat})}
\def\syst{({\rm syst})}

\newcommand{\bea}{\begin{eqnarray}}
\newcommand{\eea}{\end{eqnarray}}
\newcommand{\nn}{\nonumber\\}

\usepackage{color}
\definecolor{niceblue}{rgb}{0,0,1}
\definecolor{nicered}{rgb}{0.7,0.1,0.1}
\definecolor{nicegreen}{rgb}{0.1,0.5,.1}

\usepackage{graphicx}
\usepackage{subcaption}
\graphicspath{{figures/}{./}}

\title{A new $B$-flavour anomaly in $B_{d,s}\to K^{*0}\bar{K}^{*0}$: anatomy and interpretation}

\author[a]{Marcel Alguer\'o,}
\author[b,c,d]{Andreas Crivellin,}
\author[e]{S\'ebastien Descotes-Genon,}
\author[a]{Joaquim Matias,}
\author[e]{and Mart\'in Novoa-Brunet}

\affiliation[a]{Universitat Aut\`onoma de Barcelona, 08193 Bellaterra, Barcelona,\\
Institut de F\'{i}sica d'Altes Energies (IFAE), The Barcelona Institute of Science and Technology, Campus UAB, 08193 Bellaterra (Barcelona)}
\affiliation[b]{CERN Theory Division, CH--1211 Geneva 23, Switzerland}
\affiliation[c]{Physik-Institut, Universit\"at Z\"urich, Winterthurerstrasse 190, CH--8057 Z\"urich, Switzerland}
\affiliation[d]{Paul Scherrer Institut, CH--5232 Villigen PSI, Switzerland}
\affiliation[e]{Universit\'e Paris-Saclay, CNRS/IN2P3, IJCLab, 91405 Orsay, France}

\abstract{In the context of the recently measured non-leptonic decays $B_{d}\to K^{*0}\bar{K}^{*0}$ and $B_{s}\to K^{*0}\bar{K}^{*0}$  we analyse the anatomy of the $L_{VV}$ observable that compares  the longitudinal components of $B_s \to VV$ and $B_d \to VV$ decays. This  observable is cleaner than the longitudinal polarisation fraction as it is afflicted only at subleading order in a $1/m_b$ expansion by the theoretical uncertainties arising in the transverse components entering the polarisation fraction. Focusing on the particular case of $B_{d}\to K^{*0}\bar{K}^{*0}$ and $B_{s}\to K^{*0}\bar{K}^{*0}$, we discuss the main sources of hadronic uncertainty in the SM. We find for the SM prediction $L_{K^* \bar{K}^*}=19.5^{+9.3}_{-6.8}$,  which implies a  $2.6\sigma$ tension with respect to the most recent data, pointing to a deficit in the $b \to s$ transition of the non-leptonic decay versus the corresponding $b \to d$  transition. We  discuss possible New Physics explanations for this deviation, first at the level of the Weak Effective Theory and we identify that the two Wilson coefficients ${\cal C}_{4}$ and ${\cal C}_{8g}$  can play a central role in explaining this anomaly. Finally, we briefly explore two different simplified New Physics models which can explain the anomaly through a contribution either in ${\cal C}_{4}$ (Kaluza-Klein gluon) or in ${\cal C}_{8g}$, with a significant amount of fine tuning, but possible connections to the $b \to s \ell \ell$ anomalies.}

\emailAdd{malguero@ifae.es}
\emailAdd{andreas.crivellin@cern.ch}
\emailAdd{sebastien.descotes-genon@ijclab.in2p3.fr}
\emailAdd{matias@ifae.es}
\emailAdd{martin.novoa@ijclab.in2p3.fr}

\arxivnumber{2011.07867}

\begin{document}

\maketitle
\section{Introduction}
\label{intro}

The flavour anomalies observed in semileptonic rare B meson decays constitute  one of the most promising hints of New Physics (NP) found at LHC and $B$-Factories. Recent global analyses of the set of observables governed by the $b\to s \ell\ell$  transitions~\cite{Alguero:2019ptt,addendum}  provide a small $p$-value (1.4\%) for the Standard Model (SM), whereas simple NP hypotheses obtain a much better description of the data, with pulls up to 6.5$\sigma$ with respect to the SM (similar results are obtained in other works \cite{Alok:2019ufo,Kowalska:2019ley,DAmico:2017mtc,Aebischer:2019mlg,Datta:2019zca,Bhom:2020lmk,Biswas:2020uaq,Ciuchini:2020gvn}). A particularly promising setup combines Lepton Flavour Universality Violating (LFUV) NP together with Lepton Flavour Universal (LFU) NP, as proposed in Ref.~\cite{Alguero:2018nvb},
which improves the description of the data compared to the SM by 7.4$\sigma$ \cite{addendum} once the anomalies in $b\to c\ell\nu$ decays are included.

If NP is indeed at the origin of the anomalies in semileptonic $B$ decays, it is natural to expect signals in other observables involving $b \to s$ transitions, possibly with different realisations though sharing some common features. A natural place to explore the possible existence of these signals are non-leptonic $B$ decays. This type of  decays suffer from larger uncertainties compared to semileptonic $B$ decays and are therefore more difficult to compute with a high accuracy. In particular, branching ratios and polarisation fractions receive contributions from transverse amplitudes that suffer from large uncertainties due to power-suppressed but infrared-divergent weak annihilation and 
hard-spectator scattering~\cite{Kagan:2004uw,Beneke:2006hg}. In this sense a deviation with respect to the SM prediction in non-leptonic B decays requires one to be much more conservative regarding these uncertainties than in the case of semileptonic B decays.

In this article we will thus follow a similar strategy to the one we used in Refs.~\cite{Matias:2012xw,Descotes-Genon:2013vna} for semileptonic rare B decays and we establish a parallelism constructing  observables  in non-leptonic B decays with a limited sensitivity to hadronic uncertainties.  
This can be achieved using the $R_{sd}$ observable introduced some time ago by two of us in Ref.~\cite{DescotesGenon:2011pb}. This observable was introduced at that time 
to find NP in neutral $B$-meson mixing. However, it turns out to be particularly interesting now to find NP in the non-leptonic decay amplitudes in the light of the $b \to s \ell\ell$ anomalies for which optimized observables were introduced.

In $b\to s \ell\ell$ decays, one can build two different kinds of observables with a reduced sensitivity to hadronic uncertainties: on the one hand, angular observables from decays involving muons in the final state~\cite{DescotesGenon:2012zf,Descotes-Genon:2013vna} constructed exploiting heavy quark symmetry 
and on the other hand, ratios of branching ratios with muons versus electrons in the final state that test LFUV and where the dependence on the form factors 
cancels almost exactly in the SM~\cite{Hiller:2003js}. There are tensions in observables involving leptons of the second family (for the former) and 
between the second and the first family of leptons (for the latter). 
In this work we explore the parallel approach of using  non-leptonic B decays rather than semileptonic ones, comparing quark transitions involving quarks of the  second  and first families instead of muons and electrons.
 More specifically, we  compare transitions involving $s$-quarks and $d$-quarks to benefit from the approximate $U$-spin symmetry of the Standard Model in analogy with Lepton-Flavour Universality used to build the LFUV ratios in $b\to s\ell\ell$ decays.
The analogy has evident limitations: since both symmetries are broken by fermion mass effects, the size of the corrections is easier to 
compute or estimate for LFU (involving mainly QED) than for $U$-spin (involving QCD). However, even in the nonleptonic case it is well known that ratios of this type offer many advantages in reducing hadronic uncertainties, explaining the popularity of the ratio $\xi$ to describe neutral-meson mixing in lattice QCD and phenomenological studies. We may reach an
even better control of 
hadronic uncertainties by combining several approaches.
In Refs.~\cite{DescotesGenon:2006wc,DescotesGenon:2007qd,DescotesGenon:2011pb} two of us showed that the specific structure of 
penguin-mediated non-leptonic B-decays could lead to a better theoretical control on combinations of hadronic matrix elements
within  factorisation approaches. In the case of vector final states, it is also known that the decays into longitudinally polarised light mesons can be described more precisely than the transverse ones within these factorisation approaches, providing a further guide to build optimised observables (in analogy with the angular observables in semileptonic decays). 
Finally,  
if the $B_d$-meson decays have been studied at $B$-factories extensively, LHCb is now able to provide accurate measurements for many $B_s$-meson
decays with the possibility to assess the correlation between $B_d$ and $B_s$ mesons decaying into the same final state.

We will thus focus here on a type of observables for penguin-mediated non-leptonic decays of B mesons into two vector particles, that
we  will refer as {\it L-observables}. These correspond essentially to the $R_{sd}$ observable introduced in Ref.~\cite{DescotesGenon:2011pb} in the case of $B_{d,s} \to K^{*0} \bar{K}^{*0}$ (up to a phase space). We present here a detailed and complete anatomy of this observable in the SM, updating the SM prediction and observing an increase in the tension with the experimental measurement compared to Ref.~\cite{DescotesGenon:2011pb}. We then discuss NP explanations for the tension observed. We also point out possible improvements of the theoretical prediction of this observable.

In Sec.~\ref{sec:framework} we develop the theoretical framework that will be used to compute the $L$ observable. We put  a particular emphasis on  the sources of hadronic uncertainties coming from infrared divergences that affect mostly branching ratios and polarisations. In Sec.~\ref{sec:construction} we construct this observable and we compute it. Then using the data of the previous section we determine its experimental value and the pull. In Sec.~\ref{sec:indepNP} we explore
possible solutions in terms of NP shifts to Wilson coefficients in a model-independent EFT approach, before considering particular models illustrating the difficulty to explain this non-leptonic anomaly together with the $b \to s \ell\ell$ anomalies in Sec.~\ref{sec:NPmodels}.
We finally conclude in Sec.~\ref{sec:conclusions}. Appendices are devoted to a discussion of the weak effective theory and QCD factorisation, the semi-analyical description of relevant hadronic matrix elements, and complementary material concerning the sensitivity of $L$ to different sources of NP.
 
 \section{Theoretical framework}\label{sec:framework}
 
 \subsection{Helicity amplitudes}\label{sec:helampl}
 
We start by considering the theoretical description of $B_Q \to VV$ with $Q=d,s$. Since the initial state has spin 0, the two vector mesons must have the same helicity, leading to a description of the decay in terms of three helicity amplitudes $A^0$, $A^+$ and $A^-$. In naive factorisation one expects a hierarchy of the type:
${\bar A}^0 > {\bar A}^{-} > {\bar A}^+$ for a ${\bar B}\to VV$ decay and ${ A}^0 > { A}^{+} > { A}^-$ for a ${ B}\to VV$ decay.  This hierarchy with a dominance of longitudinal amplitudes is easy to understand by means of the V-A structure of the SM~\cite{Kagan:2004ia}. Each amplitude is suppressed with respect to the previous one by ${\cal O}(\Lambda/m_b)$ due to  helicity suppression~\cite{Kagan:2004uw}.
The longitudinal amplitude in a $b \to s$ transition is dominant as compared to the positive helicity:  the $s$ quark is produced with an helicity $-1/2$ by weak interactions (in the limit $m_s\to 0$), which is not affected by the strong interactions, then the strange quark combines with the light spectator quark to form a $V$ with a helicity which can reach $0$ or $-1$ but not $+1$.
In ${\bar A}^-$, a light-quark helicity flip is required to obtain both vector mesons with a negative helicity, whereas in ${\bar A}^+$, two helicity flips are required to reach a positive helicity for both vector mesons.
Each of these helicity flips yields a suppression by a factor ${\cal O}(\Lambda/m_b)$, as expected in naive factorisation. 
 
 \subsection{Hadronic matrix elements}

For a $\bar B_Q$ meson decaying through a $b\to q$ penguin-mediated process into a $V_1V_2$ state with a definite polarisation, the decomposition
\begin{equation}
\bar{A}_f\equiv A(\bar{B}_Q\to V_1 V_2)
  =\lambda_u^{(q)} T_q + \lambda_c^{(q)} P_q\,,
\label{dec}
\end{equation}
is always possible, with the CKM factors $\lambda_U^{(q)}=V_{Ub} V_{Uq}^*$. We denote by $T_q$ and $P_q$ the matrix elements accompanying the $\lambda_u^{(q)}$ and $\lambda_c^{(q)}$ CKM factors respectively. In the SM, $P_q$ is usually associated to  penguin topologies, whereas $T_q$ receives contributions from tree topologies (but it can also contain only penguin topologies in some decays).
As discussed above, if we consider the longitudinal polarisation, $T_q$ and $P_q$ can be computed using factorisation approaches based on a $1/m_b$ expansion (see Appendix A). In QCD factorisation~\cite{Beneke:2001ev}, $T_q$ and $P_q$ are affected by possibly large long-distance $1/m_b$-suppressed effects that will be discussed in the next section.
In the case of penguin mediated decays like $B_{(d,s)}\to K^{*0}\bar{K}^{*0}$, 
it was observed~\cite{DescotesGenon:2006wc,DescotesGenon:2007qd} that the same type of (long-distance) infrared divergences affect both $P_q$ and $T_q$, so one can construct 
\begin{equation}
\Delta_q=T_q-P_q\,, \label{delta}
\end{equation}
free from these next-to-leading-order infrared divergences.

Using the unitarity relation $\lambda_u^{(q)}+\lambda_c^{(q)}+\lambda_t^{(q)}=0$, we can write Eq.~(\ref{dec}) in terms of $\lambda_u^{(q)}$ and $\lambda_t^{(q)}$
\begin{equation}
    \label{newamp}
\bar{A}_f 
  =\lambda_u^{(q)}\, \Delta_q - \lambda_t^{(q)} P_q\,.
\end{equation}
 The weak phase in $\lambda_t^{(q)}$ is the angle $\beta_q$, defined as
\begin{equation}
    \beta_q\equiv \arg \left(- \frac{V_{tb} V_{tq}^*}{V_{cb} V_{cq}^*} \right)= \arg \left(- \frac{\lambda_t^{(q)}}{\lambda_c^{(q)}} \right)\,,
\end{equation}
whereas $\lambda_c^{(q)}$ is real to a very good approximation for both $q=d,s$,
and $\lambda_u^{(q)}=-\lambda_c^{(q)}-\lambda_t^{(q)}$.
The CP-conjugate amplitude is given by 
\begin{equation}
A_{\bar{f}}=(\lambda_u^{(q)})^* T_q + (\lambda_c^{(q)})^* P_q    =(\lambda_u^{(q)})^* \Delta_q - (\lambda_t^{(q)})^* P_q\,.
\end{equation}
If $f=V_1V_2$ is a CP-eigenstate, note that $A_{\bar{f}}$ is different from $A=A(B\to  V_1 V_2)$, even though the two types of amplitudes are related:
\begin{equation}
\bar{A}=\bar{A}_f\qquad A=\eta_f A_{\bar{f} }\,,
\end{equation}
where $\eta_f$ is the CP-parity of the final state, given for $j=0,||,\perp$ respectively as $\eta,\eta,-\eta$ where $\eta=1$ if $V_1$ is the charge conjugate of $V_2$ (this is the case for $K^{*0}\bar{K}^{*0}$).

\section{The $L$-observable for $B_Q\to K^{*0}\bar{K}^{*0}$}\label{sec:construction}
\subsection{Definition and experimental determination}

The 2019 LHCb analysis with 3 fb$^{-1}$ data measured the ratio of the untagged and time-integrated decay rates~\cite{Aaij:2019loz}
\begin{eqnarray}\label{eq:BRratioKstarKstar}
\frac{\mathcal B_{B_d\to K^{*0}\bar{K}^{*0}}}{\mathcal B_{B_s\to K^{*0}\bar{K}^{*0}}} = 0.0758 &\!\pm\!& 0.0057 \stat \pm 0.0025 \syst \nonumber \\ &\pm& 0.0016 \, \left(\frac{f_s}{f_d} \right),
\end{eqnarray} 
The longitudinal polarisation of both modes has been  measured as well. The average of $B_d\to K^{*0}\bar{K}^{*0}$ from LHCb~\cite{Aaij:2019loz} and Babar\cite{Aubert:2007xc}
\begin{eqnarray}
     f_L^{\rm LHCb}(B_d\to K^{*0}\bar{K}^{*0}) &=& 0.724 \pm 0.051 \pm 0.016, \\
     f_L^{\rm Babar}(B_d\to K^{*0}\bar{K}^{*0}) &=& 0.80^{+0.10}_{-0.12}  \pm 0.06,
\end{eqnarray}
yields
\begin{equation}
     f_L(B_d\to K^{*0}\bar{K}^{*0}) = 0.73 \pm 0.05,
\end{equation}
whereas the polarisation for the $B_s\to K^{*0}\bar{K}^{*0}$ mode is~\cite{Aaij:2019loz}:
\begin{equation}
     f_L(B_s\to K^{*0}\bar{K}^{*0}) = 0.240 \pm 0.031 \stat \pm 0.025 \syst \nn \,.
\end{equation}

Most of the experimental determinations are made assuming no direct CP-violation; however, the ones searching for CP violation found no hint in these decays~\cite{Aaij:2017wgt}.

One can notice already that the longitudinal polarisations are very different for these two modes, although they are related by $U$-spin symmetry in its most obvious form, i.e. the $d\leftrightarrow s$ exchange. In the SM, $U$-spin is broken only by the quark masses, and it is thus expected to be fairly well obeyed (up to a 20-30\% correction). We propose to define an observable that will be sensitive to this effect but with a cleaner theoretical prediction: 
\begin{equation}\label{eq:LKstKst0}
L_{V_1V_2}=\frac{{\cal B}_{b \to s}}{{\cal B}_{b \to d}}\frac{g_{b \to d} f_L^{b \to s}}{g_{b \to s} f_L^{b \to d}}=\frac{|A_0^s|^2+ |\bar A_0^s|^2}{|A_0^d|^2+ |\bar A_0^d|^2}\,,
\end{equation}
where 
${\cal B}_{b \to q}$ ($f_L^{b \to q}$) refers to the branching ratio (longitudinal polarisation) of the $\bar{B}_Q \to V_1V_2$ decay governed by a $b \to q$ transition. $A_0^q$ and $\bar{A}_0^q$ are the 
amplitudes for the $B_Q$ and $\bar{B}_Q$ decays governed by $b \to q$ with final vector mesons being polarised longitudinally and
\begin{equation}
g_{b\to q}=\omega \sqrt{\left[M_{B_Q}^2-\Sigma_{V_1V_2}\right]\left[M_{B_Q}^2-\Delta_{V_1V_2}\right]}\,,
\end{equation}
 stands for the phase space factor involved in the corresponding branching ratio, 
with $\omega=\tau_{B_Q}/(16\pi M_{B_Q}^3)$,
$\Sigma_{ab}=(m_a+m_b)^2$ and $\Delta_{ab}=(m_a-m_b)^2$
and all quantities are CP-averaged. 
 
This observable is defined such that the dependence on the troublesome transverse (parallel and perpendicular) amplitudes entering the branching ratio and longitudinal polarisation fraction cancel and it is close to the observable $R_{sd}$ for the case of $B_{d,s} \to K^{*0} \bar{K}^{*0}$ up to a phase space factor~\cite{DescotesGenon:2011pb}.

Being purely sensitive to the longitudinal amplitudes, $L$ is less affected by the hadronic uncertainties which impact the transverse polarisation amplitudes significantly and which are difficult to estimate within QCD Factorisation (QCDF) or other approaches based on a $1/m_b$ expansion. The choice of this observable thus avoids the difficulties encountered in the interpretation of low longitudinal polarisation fractions observed in some non-leptonic modes~\cite{Kagan:2004uw}. 
In this article we will focus on:
\begin{equation}\label{eq:LKstKst}
L_{K^*\bar{K}^*}=\frac{{\cal B}_{B_s \to K^{*0} {\bar K^{*0}}}}{{\cal B}_{B_d \to K^{*0} {\bar K^{*0}}}}\frac{g_{b\to d} f_L^{B_s}}{g_{b\to s} f_L^{B_d}}=\frac{|A_0^s|^2+ |\bar A_0^s|^2}{|A_0^d|^2+ |\bar A_0^d|^2}\,,
\end{equation} 
where the spectator quark $Q$ of the initial $b$-flavoured meson and the quark $q$ from the $b \to q$ transition coincide.

In the definition of $L_{K^*\bar{K}^*}$ and its connection with the longitudinal amplitudes $|A_0^q|^2$ in Eq.~(\ref{eq:LKstKst}), we have not included the effect of $B_s$-meson mixing
that arises in branching ratios when measured at  hadronic machines. This effect of time integration at hadronic machines
generates a correction of ${\cal O}(\Delta \Gamma/(2\Gamma))$ discussed in Refs.~\cite{DescotesGenon:2011pb,DeBruyn:2012wj}, which would multiply the last term  in Eq.~(\ref{eq:LKstKst}) by:
\begin{equation} \label{eq:timeintegration}
    \frac{1+A_{\Delta\Gamma}^s y_s}{1+A_{\Delta\Gamma}^d y_d}
    \frac{1-y_d^2}{1-y_s^2},
\end{equation}
where $y_q=\Delta \Gamma_{B_q}/(2\Gamma_{B_q})$ is well measured ($y_d$ is negligible and $y_s\simeq 0.065$) and  the asymmetries
$-1\leq A_{\Delta\Gamma}^q\leq 1$ combining $CP$ violation in mixing and decay are difficult to estimate theoretically, leading to a correction of at most 7\%.

Since we use  the LHCb measurement Eq.~(\ref{eq:BRratioKstarKstar}) and since there are other sources of (theoretical and experimental) uncertainties, we treat Eq.~(\ref{eq:timeintegration}) as a
systematic uncertainty of 7\% combined in quadrature with the other uncertainties, leading to the experimental value:
\begin{equation}\label{eq:expL}
   {\rm Exp}:\qquad L_{K^*\bar{K}^*}=4.43\pm 0.92. 
\end{equation}

\subsection{Theoretical prediction in the SM and comparison with data}
On the theory side, we have
\begin{eqnarray}
A_0^q & = & (\lambda_c^{(q)*}+\lambda_u^{(q)*})\left[ P_{q} + (\alpha^{q})^*
 \Delta_{q} \right],\\
\bar{A}_0^q & = & (\lambda_c^{(q)}+\lambda_u^{(q)})\left[ P_{q} + \alpha^{q} \Delta_{q} \right],
\label{eqsP}
\end{eqnarray}
where $\alpha^q=\lambda_u^q/(\lambda_c^q+\lambda_u^q)$. We thus get
  \begin{equation}\label{eq:LKstKstDeltaP}
\!\! L_{K^*\bar{K}^*}=  \kappa \left|\frac{P_s}{P_d}\right|^2 
 \left[\frac{1+\left|\alpha^s\right|^2\left|\frac{\Delta_s}{P_s}\right|^2
 + 2 {\rm Re} \left( \frac{ \Delta_s}{P_s}\right) {\rm Re}(\alpha^s) 
 }{1+\left|\alpha^d\right|^2\left|\frac{\Delta_d}{P_d}\right|^2
  + 2 {\rm Re} \left( \frac{ \Delta_d}{P_d}\right) {\rm Re}(\alpha^d)} \right]\,,
 \end{equation}
 with the combinations of CKM factors (estimated using the
 summer 2019 CKMfitter update~\cite{Charles:2004jd,Charles:2016qtt,Koppenburg:2017mad}
 (see Table~\ref{tab:inputs}):
  \begin{eqnarray}
 \alpha^d&=&(-0.0136^{+0.0095}_{-0.0096}) +  i (0.4181^{+0.0085}_{-0.0064}), \\
 \alpha^s&=&(0.00863^{+0.00040}_{-0.00036})+i (-0.01829^{+0.00037}_{-0.00042}),\\
 \kappa&=&\left|\frac{\lambda_c^{s}+\lambda_u^{s}}{\lambda_c^{d}+\lambda_u^{d}}\right|^2 = 22.92^{+0.52}_{-0.30}. \label{eq:kappa} \end{eqnarray}
 
From QCD factorisation and the discussion in Sec.~\ref{sec:framework}, we have
 \begin{eqnarray}
 \frac{\Delta_d}{P_d}&=&
 (-0.16\pm 0.15) + (0.23\pm 0.20) i,
 \nonumber \\
\frac{\Delta_s}{P_s}&=& (-0.15 \pm 0.22) + (0.23\pm 0.25) i,
 \end{eqnarray}
 so that the brackets in Eq.~(\ref{eq:LKstKstDeltaP}) are very close to 1, with the main uncertainty of 1\% from the term proportional to $|\alpha^d|^2$  (which will be included in the theoretical uncertainties below). 
 The leading uncertainty in the theoretical evaluation of $L_{K^*\bar{K}^*}$ comes thus from the ratio $|P_s/P_d|$, which we can attempt to estimate in different ways.
 A naive $SU(3)$ approach would consist in assuming
 \begin{equation}\label{eq:Pratio1}
     {\rm naive}\ SU(3): \left|\frac{P_s}{P_d}\right|=1\pm 0.3\,,
 \end{equation}
while a naive factorisation approach would rather yield
  \begin{equation}\label{eq:Pratio2}
     {\rm fact}\ SU(3): \left|\frac{P_s}{P_d}\right|=f=0.91^{+0.20}_{-0.17}\,,
 \end{equation}
 where the $SU(3)$-breaking ratio related to the form factors of interest is given by
  \begin{equation}
    f=\frac{ A^s_{K^*\bar{K^*}}}{A^d_{K^*\bar{K^*}}}
     =\frac{m_{B_s}^2  A^{B_s \to {K^*}}_0(0)}{m_{B_d}^2  A^{B_d \to {K^*}}_0(0)}\,,
 \end{equation}
 and we used the values of Ref.~\cite{Straub:2015ica} for the form factors to estimate $f$.
 A last possibility amounts to using QCD factorisation. Using the same inputs as before, we obtain
 \begin{equation}\label{eq:Pratio3}
   {\rm QCD\ fact}:  \left|\frac{P_s}{P_d}\right|=0.92^{+0.20}_{-0.18}\,. 
 \end{equation} 

 The QCD factorisation-based prediction follows the theoretical computations of the different contributions to the amplitudes from Refs.~\cite{Beneke:2006hg,Beneke:2003zv}. The numerical values of the input parameters used are updated with respect to the ones in Ref.~\cite{Beneke:2003zv} and  can be found in Table~\ref{tab:inputs} of Appendix~\ref{app:WET}.
 
  \begin{table}[h]
	\begin{center}
\begin{tabular}{|c|c|c|}
\hline
Observable   & 1$\sigma$ & 2$\sigma$  \\
\hline
$L_{K^*\bar{K}^*}$ & $[12.7, 28.8]$ & $[7.5, 43]$ \\
\hline
\end{tabular}
	\caption{1$\sigma$ and 2$\sigma$ confidence intervals for the SM prediction of $L_{K^*\bar{K}^*}$ within QCD factorisation.}
		\label{tab:confidence}
    \end{center}
\end{table}
 
Hard-gluon exchanges with the spectator quark and weak annihilation feature $1/m_b$-suppressed contributions exhibiting infrared divergences related to the endpoint of the meson light-cone distribution amplitudes. These divergences are parametrised in the same manner as in Ref.~\cite{Beneke:2003zv}, involving two contributions $X_H$ and $X_A$ treated as universal
 for all channels:
  \begin{equation}
     X_{H,A}=(1+\rho_{H,A}e^{i\varphi_{H,A}})\ln{\left(\frac{m_B}{\Lambda_{h}}\right)}\,.
 \end{equation}
 We take $\rho_{H,A}\in[0,1]$ and $\varphi_{H,A}\in[0,2\pi]$ with flat distributions.
 This translates into assigning a $100\%$ uncertainty to the magnitude
 of such corrections.
 
 We propagate the uncertainties by varying each input (given in Tab.~\ref{tab:inputs}) entering the penguin ratios in Eqs.~(\ref{eq:Pratio1}), (\ref{eq:Pratio2}) and (\ref{eq:Pratio3}) and the CKM contribution $\kappa$ following Eq.~(\ref{eq:kappa}), using Gaussian distributions. We determine then the distribution of $L$ in each case, leading to the $1 \sigma$ ranges:
  \begin{eqnarray}\label{eq:tension1}
     {\rm naive}\ SU(3):    L_{K^*\bar{K}^*}=& 23^{+16}_{-12}  \qquad \hspace{3mm} 1.9\sigma\,,\\ 
   {\rm fact}\ SU(3):    L_{K^*\bar{K}^*}=& 19.2^{+9.3}_{-6.5} \qquad 3.0 \sigma\,,\\ 
  {\rm QCD\ fact}:    L_{K^*\bar{K}^*}=& 19.5^{+9.3}_{-6.8} \qquad 2.6 \sigma\,, 
  \label{eq:tension3}
  \end{eqnarray}
  where we put the level of discrepancy with experiment, in units of $\sigma$. We stress
  that these discrepancies are obtained using the whole distribution for $L$
  and not just the $1\sigma$ confidence intervals in the Gaussian approximation (see Tab.~\ref{tab:confidence} for the 1 and $2\sigma$ confidence intervals).
  In Tab.~\ref{tab:errorbudget} we present the error budget for $L_{K^{*}\bar{K}^{*}}$ in the SM. The comparison with the error budget of $|P_{d,s}|^2$ shows that the impact of $X_A$ ($X_H$)  is
  reduced from 18\% (2\%) in $|P_{d,s}|^2$ to 4\% (0.2\%) in $L_{K^*\bar{K}^*}$. A similar reduction is observed for other inputs such as $f_{K^*}$,  showing the benefit of defining the ratio $L_{K^*\bar{K}^*}$.
  It also indicates that the accuracy of the theoretical prediction of $L_{K^*\bar{K}^*}$ could be improved significantly by determining the correlations among the relevant $B\to K^*$ form factors
  in order to compute the associated $SU(3)$ breaking. Moreover, the impact of the weak annihilation and hard-scattering divergences on the uncertainty is subdominant and would not be affected strongly by using a different approach for these power-suppressed infrared divergences.
 
From the comparison of the SM predictions Eqs.~(\ref{eq:tension1})-(\ref{eq:tension3}) with the experimental result in Eq.~(\ref{eq:expL}), we see that all our theoretical estimates point towards a deficit in the $b \to s$ transition compared to the $b \to d$ one for these penguin-mediated modes, in analogy with the deficit observed in semileptonic decays to muons versus the decay to electrons in $b\to s\ell\ell$ decays.

      \begin{table}[h]
	\begin{center}
$$
\begin{array}{|c||c|c|c|}
\cline{2-4}
\multicolumn{1}{c|}{}&\multicolumn{3}{c|}{\text{Relative Error}}\\ 
\hline
\text{Input} & L_{K^{*}\bar{K}^{*}} & |P_s|^2 & |P_d|^2 \\ \hline\hline
 f_{K^*} & (-0.1\%,+0.1\%) & (-6.8\%,+7.1\%) &   (-6.8\%,+7\%) \\\hline
 A^{B_d}_{0} & (-22\%,+32\%) & - & (-24\%,+28\%) \\ \hline
 A^{B_s}_{0} & (-28\%,+33\%) & (-28\%,+33\%) & - \\\hline
 \lambda_{B_d} & (-0.6\%,+0.2\%) & (-4.6\%,+2.1\%) &
   (-4.1\%,+1.9\%) \\\hline
 \alpha_2^{K^*} & (-0.1\%,+0.1\%) & (-3.6\%,+3.7\%) &
   (-3.6\%,+3.6\%) \\\hline
 X_H & (-0.2\%,+0.2\%) & (-1.8\%,+1.8\%) &
   (-1.6\%,+1.6\%) \\\hline
 X_A & (-4.3\%,+4.4\%) & (-17\%,+19\%) & (-13\%,+14\%)
   \\\hline
 \kappa  & (-1.4\%,+2.2\%) & - & - \\\hline
 \text{Others} & (-1.3\%,+1.1\%) & (-2.7\%,+2.5\%) &
   (-1.6\%,+1.6\%) \\\hline
\end{array}
$$

	\caption{Error budget of $L_{K^{*}\bar{K}^{*}}$ and $|P_{d,s}|^2$. The relative error of each theoretical input is obtained by varying
	them individually. The main sources of uncertainty are the form factors, followed by weak annihilation at a significantly smaller level.}
		\label{tab:errorbudget}
    \end{center}
\end{table}

\section{Model-independent  NP analysis}\label{sec:indepNP}
  
 Even though the deviation in $L_{K^*\bar{K}^*}$ is not yet at the level of a troublesome discrepancy with the SM, its potential connection with other $B$-flavour anomalies makes it interesting to investigate it further in terms of possible $SU(3)$-breaking NP contributions.
We may explore in a model-independent way how to explain this anomaly via contributions only to the Wilson coefficients of the $b \to s$ transition, while keeping the corresponding $b \to d$ SM-like (or with  opposite NP contributions). 

This can be performed by using  the weak effective theory, whose basis within the SM we recall in Eq.~(\ref{eq:wet}) of App.~\ref{app:WET}. Note that in the presence of generic NP, the basis of operators  must be extended since we expect this NP contribution to couple with  different strength to different flavours (and in particular to $d$ and $s$ quarks), there is no a priori reason for it to yield ``strong'' and ``electroweak'' penguin operators 
with sums over all quark flavours following the same pattern as in the SM~\cite{Grossman:1999av}.
 
However, for simplicity, and in parallel with the results of the global fits for NP in $b\to s\ell\ell$ decays favouring SM operators or chirally-flipped versions of it, we consider here only NP entering the Wilson coefficients associated with the SM operators $Q_i$ or the chirally-flipped ones $\tilde{Q}_i$ as defined in Ref.~\cite{Kagan:2004ia} by exchanging $V-A$ and $V+A$ in all quark bilinears constituting the operators. These right-handed currents would modify the longitudinal amplitude by adding contributions that are functions
 of ${\cal C}_i^{\rm NP}-\tilde{{\cal C}}_i$ (where $\tilde{\cal C}_i$ is the coefficient of the chirally-flipped operator) leading to the structure
$ A_0[{\cal C}_i^{\rm SM}]+ A_0[{\cal C}_i^{\rm NP}-\tilde{{\cal C}}_i]$. In practice this means that the NP contribution to each coefficient entering the longitudinal amplitude should be interpreted  as stemming not only from the standard operators but also from the chirally flipped ones (with an opposite sign).
 
 We consider the sensitivity of $L_{K^*\bar{K}^*}$ on each Wilson coefficient.
We want to determine if there is a dominant operator
 that can naturally explain the low experimental value of $L_{K^*\bar{K}^*}$, as it happens for $b \to s \ell\ell$ with ${O}_9$. 
 We assume that NP enters as described above with the further requirement that there are no additional NP phases, leading to real-valued  Wilson coefficients. 
  We can then compute the hadronic matrix elements within QCD factorisation exactly like in the SM.
In Appendix~\ref{app:semianalytical} we provide semi-analytical expressions for $P_d$ and $P_s$, needed to compute $L_{K^*\bar{K}^*}$ in terms of Wilson coefficients. We provide the explicit dependence on the infrared divergences $X_A$ and $X_H$ although their numerical impact on the uncertainty is limited. Let us note in passing that the quantity $\Delta_q$ is still protected from infrared divergences in this NP extension:  the structure of the longitudinal hadronic amplitudes $T$ and $P$ is unchanged, and only the numerical values of Wilson coefficients are modified compared to the SM (the protection of $\Delta$ from infrared divergences would not necessarily hold in more general NP extensions). 

Considering the sensitivity of $L_{K^*\bar{K}^*}$ on each Wilson coefficient of the weak effective theory individually, we can determine the coefficients  where a limited NP contribution would be sufficient to explain the discrepancy observed. We thus identify three dominant coefficients: ${\cal C}_{1q}^{c}$, ${\cal C}_{4q}$ and ${\cal C}_{8gq}^{\rm eff}$ (see Fig.~\ref{fig:C1sC4sC8gs} and Fig.~\ref{fig:s} in Appendix~\ref{app:plots}).  The strong dependence on these coefficients with respect to the others can be seen already in the explicit form of $P_{d,s}$:
\begin{eqnarray}
\nonumber
    P_s&=&(1.98-5.04i)+(2.37-1.65i){\cal C}_{1s}^{c,{\rm NP}}+ (9.98+148.76i) {\cal C}_{4s}^{\rm NP}  - 7.98i {\cal C}_{8gs}^{\rm eff,NP} +\ldots \, \, \label{eq:PsCi}\\ \nonumber
    P_d&=&(2.17-5.49i)+(2.60-1.80i){\cal C}_{1d}^{c,{\rm NP}} + (10.95+161.74i) {\cal C}_{4d}^{\rm NP} - 8.76i {\cal C}_{8gd}^{\rm eff,NP} +\ldots \, \, \, \, \label{eq:PdCi}
\end{eqnarray}

which translates into a dominant contribution for $L_{K^*\bar{K}^*}$ as well. 

\begin{figure}[b]
\includegraphics[width=0.5\textwidth]{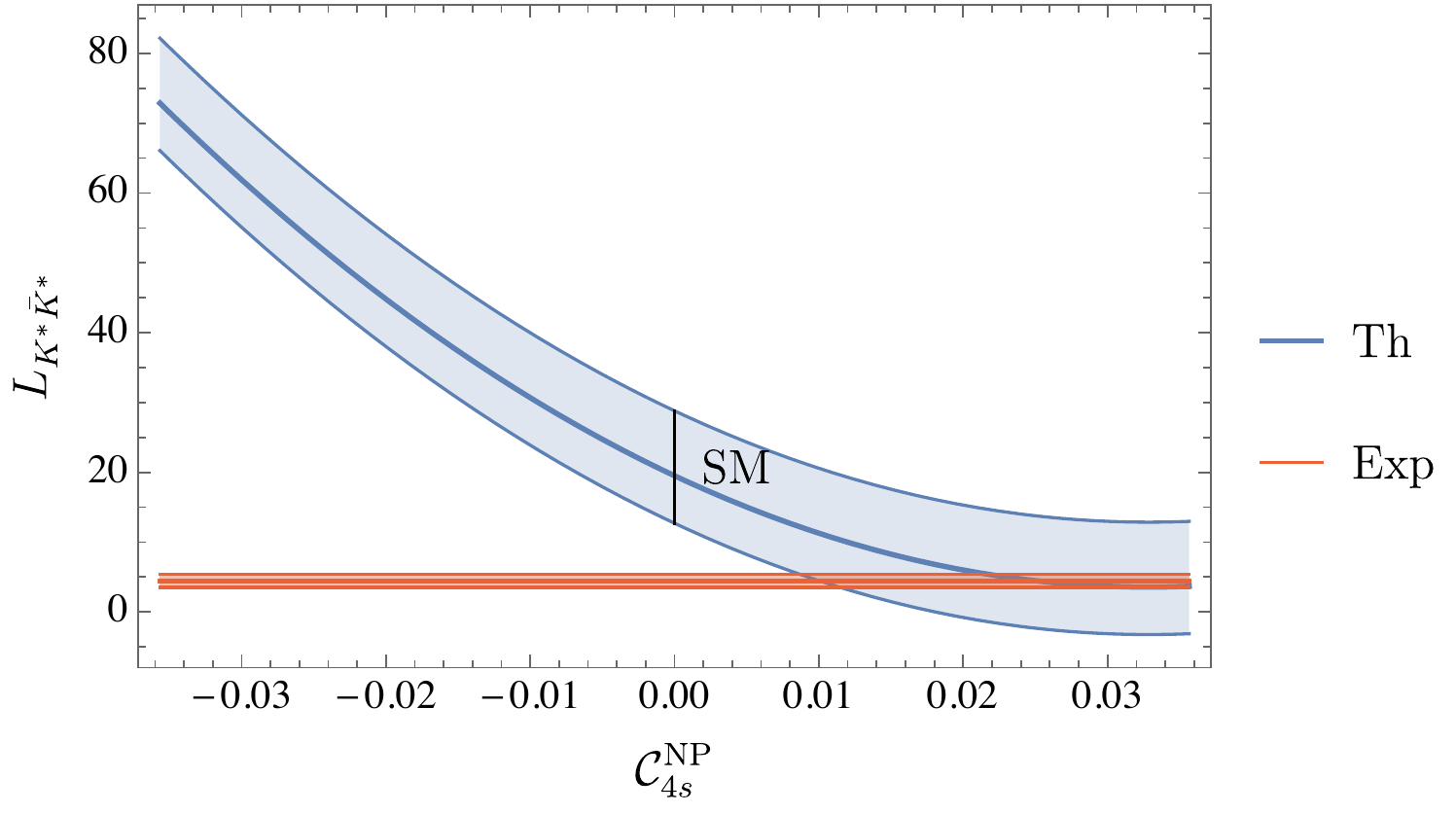}
\includegraphics[width=0.5\textwidth]{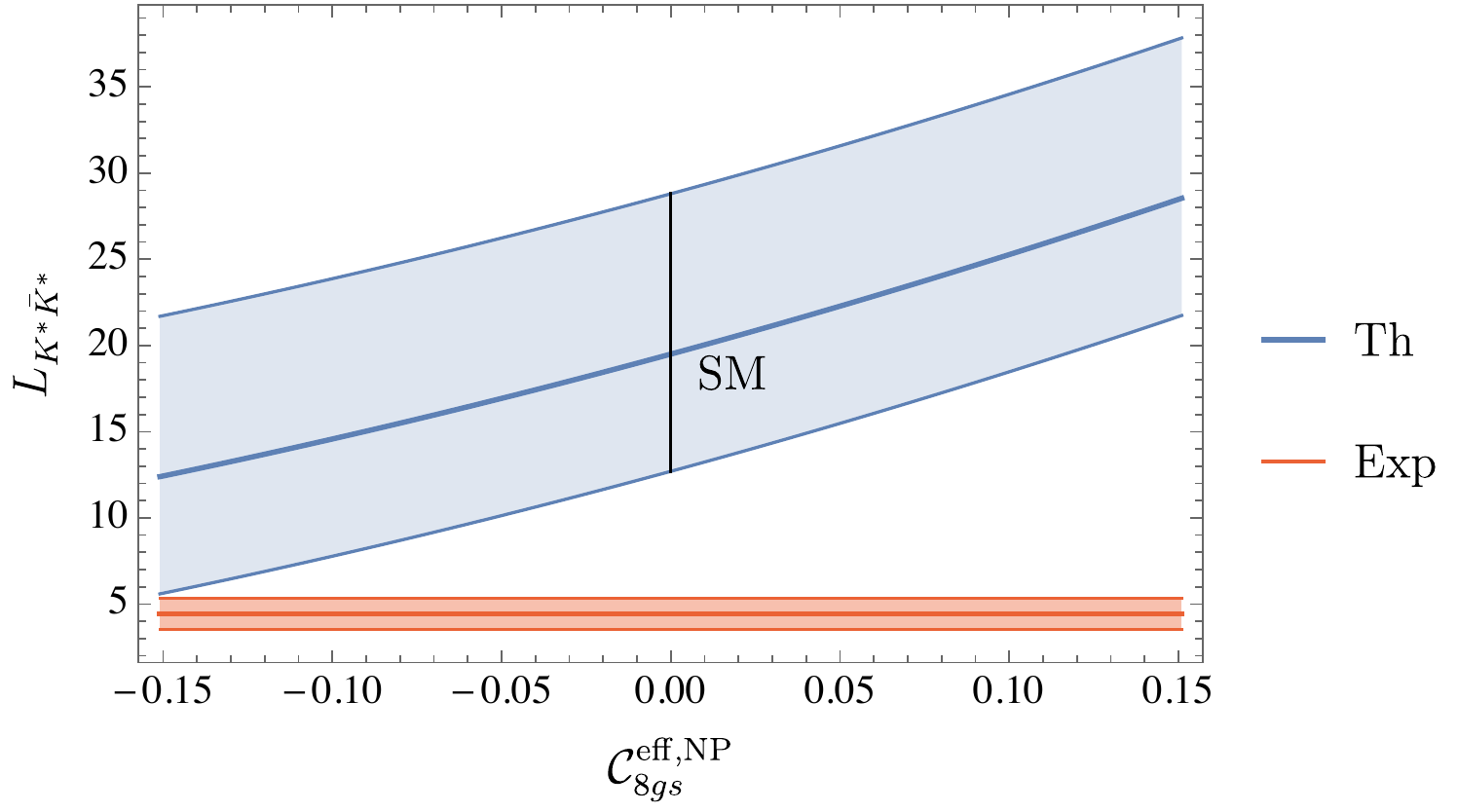}
\caption{\label{fig:C1sC4sC8gs} The tension between the theoretical prediction (blue) and the experimental value (orange) is reduced below $1\sigma$ for ${\cal C}^{\rm NP}_{4s}\simeq 0.25{\cal C}^{\rm SM}_{4s}$ (upper plot) or ${\cal C}^{\rm eff,NP}_{8gs}\simeq -{\cal C}^{\rm eff,SM}_{8gs}$ (lower plot). The predictions are given for ${\cal C}^{\rm NP}_{4s}$ and ${\cal C}^{\rm eff,NP}_{8gs}$  for a range corresponding to 100\% of their respective SM values. The plots for the remaining Wilson coefficients can be found in Appendix~\ref{app:plots}.}
\end{figure}

\begin{figure}[b]
\centering
\includegraphics[width=0.5\textwidth]{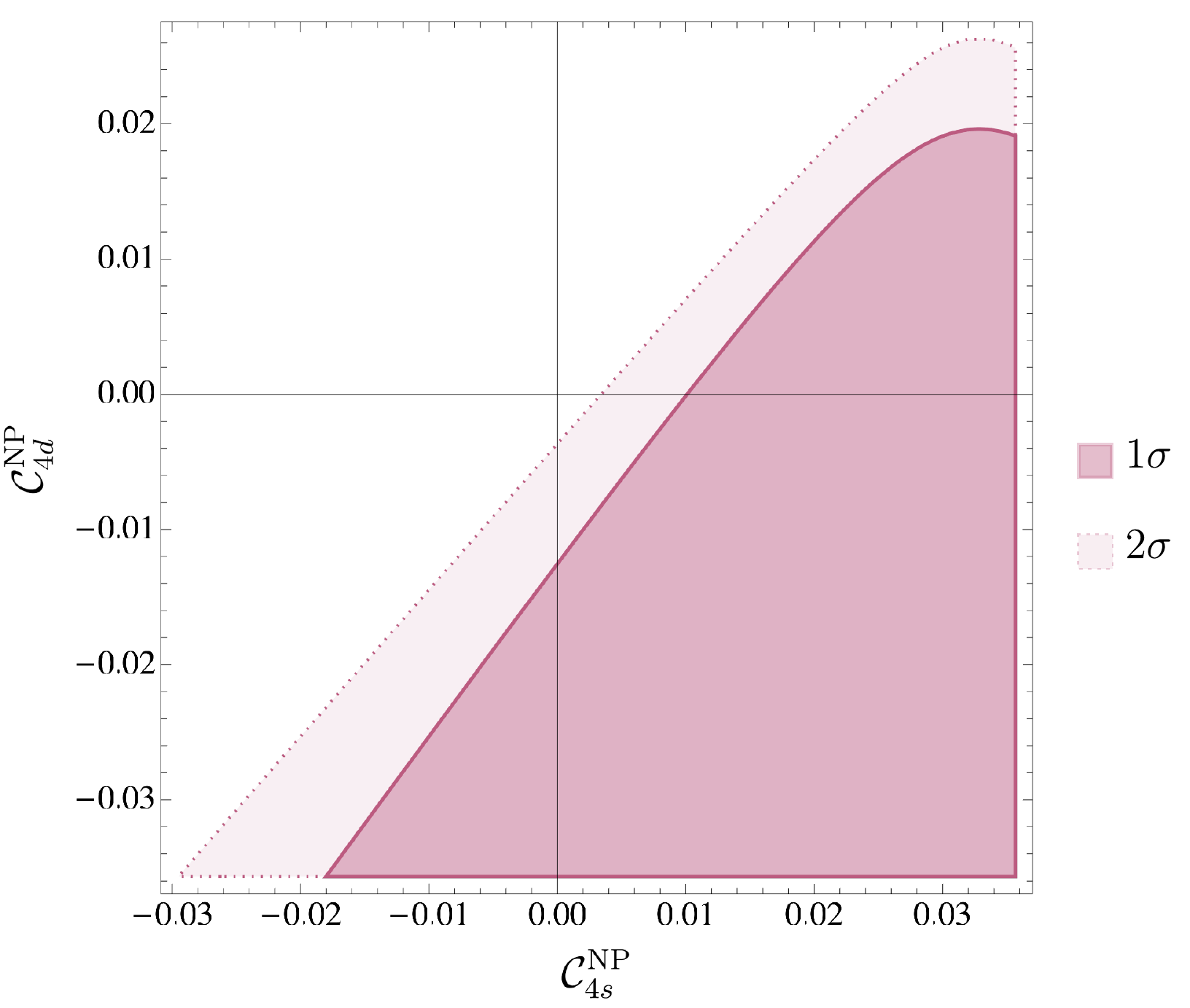}
\caption{\label{fig:C4svsC4d} 1$\sigma$ and $2\sigma$ CL regions from $L_{K^{*}\bar{K}^{*}}$ allowing NP contributions to both ${\cal C}_{4s}$ and ${\cal C}_{4d}$.}
\end{figure}

The reason behind this  strong dependence on these coefficients can be understood in the following way. Let us consider a penguin-mediated decay, so that the SM tree-level operator ${\cal C}_{1s}^{c}$ contributes through a closed $c\bar{c}$ loop to the decay, putting its contribution at the same level as the ``strong" penguin operators $i=3\ldots 6$ in the SM. A very similar contribution at the level of the underlying SM diagrams comes thus from both ${\cal C}_{1s}^{c}$ and ${\cal C}_{4s}$, as can be seen from the $V-A$ structure of the operators (this is also the case for ${\cal C}_{8gs}^{\rm eff}$ with the emission of a gluon coupling to a $q\bar{q}$ pair). The effect of the diagrams is similar in the SM, but the separation between long and short distances in the weak effective theory yields ${\cal C}_{4s}$ and ${\cal C}_{8gs}^{\rm eff}$ much smaller than ${\cal C}_{1s}^{c}$, which must be compensated by larger weights in Eqs.~(\ref{eq:PsCi}) and (\ref{eq:PdCi}).
The other penguin operators are suppressed either because of colour suppression (${\cal C}_3$, thus associated with $1/N_c$ factors in the QCD factorisation formula) or helicity suppression  (${\cal C}_5$ and ${\cal C}_6$, which yield a vanishing contribution
in the naive factorisation approach as they must be Fierzed into (pseudo)scalar operators with vanishing matrix elements). In the SM, the ``electroweak'' penguins $i=7\ldots 10$
are suppressed.  Their contributions might be very significantly enhanced by NP which would not require such an electromagnetic suppression, although it would be difficult to obtain then ``electroweak'' operators at the $m_b$-scale since they involve explicitly the quark electric charges.  If we nevertheless allowed for such very large contributions for the electroweak part (which we will discard in the following), the same argument would apply as in the case of the ``strong'' penguins, so that the leading contribution from the ``electroweak'' penguins would be ${\cal C}_{10q}$.

As can be seen in Fig.~\ref{fig:s}, the  coefficient ${\cal C}_{1s}^{ c}$  requires a very large NP contribution w.r.t. the SM of order 60\% to reduce this discrepancy at $1\sigma$. We will not pursue the possibility of  a contribution to ${\cal C}_{1q}^c$, as the size of the effect being so large at an absolute scale is in conflict with recent analyses of the global constraints on this coefficient~\cite{Lenz:2019lvd} that suggest that the room for NP contributions is of ${\cal O}(10\%)$ of the SM. Dijet angular distributions~\cite{Sirunyan:2018wcm}, together with flavour bounds following from $SU(2)_L$ gauge invariance, suggest bounds which are even tighter. 

The penguin coefficient ${\cal C}_{4s}$ requires a NP contribution of order 25\%  (which is incidentally similar to the NP contribution needed in ${\cal C}_9$ for $b \to s \mu\mu$) in order to reduce the discrepancy in $L_{K^*\bar{K}^*}$ at $1\sigma$. The NP contribution needed is thus quite large but not significantly constrained from other non-leptonic decays where many other coefficients enter~\cite{Beneke:2001ev}. 

Finally, ${\cal C}_{8gs}^{\rm eff}$ would require a NP contribution of order 100\% of the SM in order to obtain a similar reduction of the discrepancy. Although it might seem a large contribution,  it is actually very difficult to obtain a precise bound on this effective coefficient which combines ${\cal C}_{8g s}$ with some Wils0on coefficients of four-quark operators (see Appendix~\ref{app:WET}). Due to QCD loop effects, the constraint from $b\to s\gamma$ is actually on a linear combination of the Wilson coefficients  ${\cal C}_{7\gamma s}^{\rm eff}$ and ${\cal C}_{8gs}^{\rm eff}$ at the scale $\mu_b$ \cite{Misiak:2020vlo}. Therefore, an effect in ${\cal C}_{8g s}^{\rm eff}$ can always be cancelled by an effect in ${\cal C}_{7\gamma s}^{\rm eff}$
so that the experimental bound from $b \to s \gamma$ is obeyed (the same is also true for $b\to d\gamma$~\cite{Crivellin:2011ba}).
Even without such a cancellation from ${\cal C}_{7\gamma s}^{\rm eff}$, the current measurements can accommodate a NP contribution to ${\cal C}_{8g s}^{\rm eff}$ of the order of the SM.
Another more direct bound on ${\cal C}_{8gs}^{\rm eff}$ is provided by the $b\to sg$ contribution to inclusive non-leptonic charmless decays. The current bound on the $b\to sg$ branching ratio in Ref.~\cite{Zyla:2020zbs} is at the level of 6.8\%, whereas the SM contribution~\cite{Greub:2000sy} is estimated at the level of 0.5\%, leaving room for a NP contribution to ${\cal C}_{8gs}^{\rm eff}$ up to three times as large as the SM one.

Naturally, in each case, if we allow for NP in both ${\cal C}_{is}$ and ${\cal C}_{id}$, we may get the same reduction of the discrepancy by assigning half of the NP contribution (with opposite signs) to both coefficients, as illustrated for ${\cal C}_4$ in Fig.~\ref{fig:C4svsC4d}. Thus, allowing NP in $b\to d$ transitions in addition to $b\to s$ transitions requires smaller NP contributions in each type of transition, and allows one to evade some of the bounds discussed above as they applied only to $b\to s$ transitions (e.g. ${\cal C}_{8gs}$). ${\cal C}_{8gd}$ is constrained from $b\to d\gamma$.

  \section{Simplified NP models}\label{sec:NPmodels}
   
 Our model-independent analysis showed that $L_{K^*\bar{K}^*}$ is mostly sensitive to
 colour-octet operators and to a lesser extent to the chromomagnetic operator. In the following, we will consider NP models able to generate such contributions, and for concreteness, present the formula for the case of $b\to s$ transitions.
 
 Concerning ${\cal C}_{4s}$, it is natural to search for a tree-level explanation in terms of NP and a massive $SU(3)_c$ octet vector particle, i.e. a Kaluza-Klein (KK) gluon, also called axi-gluon, comes naturally to mind. We parametrise its couplings to down quarks {of different flavours} as
 \begin{equation} \label{lagrangian}
 {\cal L} = \Delta _{sb}^L\bar s{\gamma ^\mu }{P_L}{T^a}bG_\mu ^a + \Delta _{sb}^R\bar s{\gamma ^\mu }{P_R}{T^a}bG_\mu^a\,.
 \end{equation}
 with $\Delta_{sb}^{L,R}$ assumed real.  We also define from Eq. (\ref{lagrangian}) analogous flavour diagonal couplings which we will denote as $\Delta_{qq}^{L,R}$.
 
\begin{figure}
\centering
\includegraphics[width=0.5\textwidth]{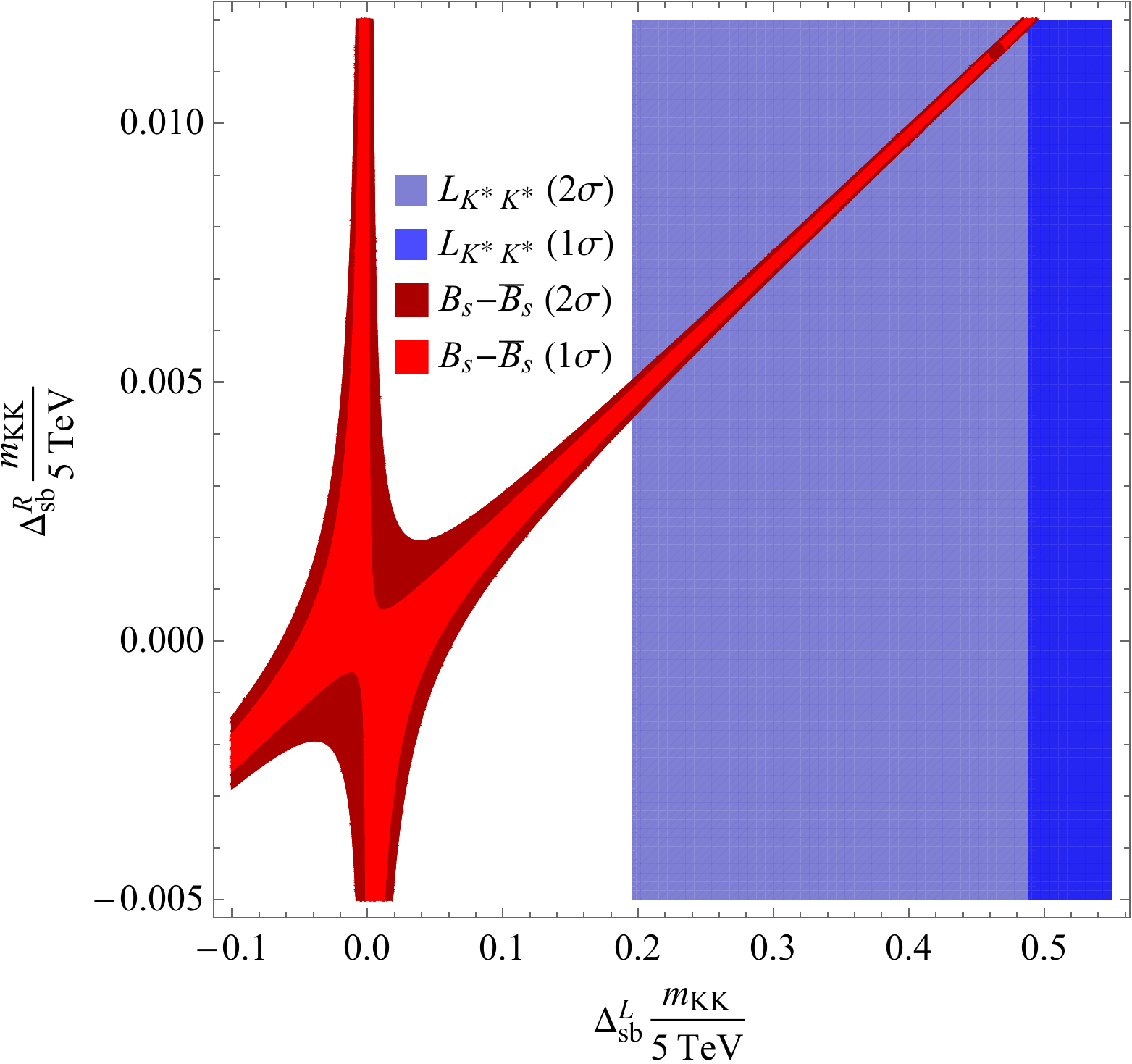}
\caption{\label{KKgluon} Preferred regions from $B_s-\bar B_s$ mixing (red) and $L_{K^*\bar{K}^*}$ (blue) for $\Delta_{qq}^R=0$ and the maximal value of $\Delta_{qq}^L$ compatible with LHC searches assuming real couplings. Note that explaining $L_{K^*\bar{K}^*}$ requires some fine-tuning in $\Delta^L_{sb}$ vs $\Delta^R_{sb}$.}
\end{figure}

We may consider the constraints from neutral-meson mixing through the effective Hamiltonian of Ref.~\cite{Becirevic:2001jj}  
 \begin{eqnarray}
 \nonumber
H_{eff}^{\Delta F = 2} &=& \sum\limits_{j = 1}^5 {{{\cal C}_j^{B_s\bar B_s}}} {\mkern 1mu} {O_j^{B_s\bar B_s}} + \sum\limits_{j = 1}^3 {{{\tilde {\cal C}}_j^{B_s\bar B_s}}} {\mkern 1mu} {{\tilde O}_j^{B_s\bar B_s}}\,,\\ 
{O_1^{B_s\bar B_s}}{\mkern 1mu}  &=& \left[ {{{\bar s}_\alpha }{\gamma ^\mu }{P_L}{b_\alpha }} \right]\left[ {{{\bar s}_\beta }{\gamma_\mu }{P_L}{b_\beta }} \right]\,,\\
{O_4^{B_s\bar B_s}}{\mkern 1mu} &=& \left[ {{{\bar s}_\alpha }{P_L}{b_\alpha }} \right]\left[ {{{\bar s}_\beta }{P_R}{b_\beta }} \right]\,,\\
{O_5^{B_s\bar B_s}}{\mkern 1mu} &=& \left[ {{{\bar s}_\alpha }{P_L}{b_\beta }} \right]{\mkern 1mu} \left[ {{{\bar s}_\beta }{P_R}{b_\alpha }} \right]\,,
\end{eqnarray}
where {only the operators relevant for the discussion are displayed} and where
the operators with a tilde are obtained by exchanging the {chirality projectors $P_L$ and $P_R$}. We get the matching contributions 
 \begin{align}
 {\cal C}_1^{B_s\bar B_s} &= \frac{1}{2m_{KK}^2}\left( {\Delta _{sb}^L} \right)^2\frac{1}{2}\left( {1 - \frac{1}{N_C}} \right)\,,\\
 \tilde {\cal C}_1^{B_s\bar B_s} &= \frac{1}{{2m_{KK}^2}}{{\left( {\Delta _{sb}^R} \right)}^2}\frac{1}{2}\left( {1 - \frac{1}{N_C}} \right)\,,\\
 {\cal C}_4^{B_s\bar B_s} &=  - \frac{1}{{m_{KK}^2}}\Delta _{sb}^L\Delta _{sb}^R\,,\\
 {\cal C}_5^{B_s\bar B_s} &= \frac{1}{N_Cm_{KK}^2}\Delta _{sb}^L\Delta _{sb}^R\,,
 \end{align}
where $m_{KK}$ is the mass of the KK gluon. Using the two-loop Renormalisation Group Equations of Refs.~\cite{Ciuchini:1997bw,Buras:2000if} and the bag factors of Ref.~\cite{Aoki:2019cca} this translates to 
  \begin{align}
\frac{\Delta M_{B_s}^{\rm NP}}{\Delta M_{B_s}^{\rm SM}} \times 10^{-10}&= \left(1.1 ({\cal C}_1^{B_s\bar B_s}+\tilde{{\cal C}}_1 ^{{B_s\bar B_s}})
+8.4 {\cal C}_4^{B_s\bar B_s}+3.1 {\cal C}_5^{B_s\bar B_s}\right){\rm GeV}^2\,,
 \end{align}  
 for a NP scale around 5 TeV. This has to be compared with the outcome of global fits
 allowing for NP in mixing~\cite{Bona:2007vi,Charles:2020dfl}, favouring a value slightly above 1 for the ratio ${\Delta M_{B_s}^{\rm exp}}/{\Delta M_{B_s}^{\rm SM}}$. 
 Encompassing the results  obtained from these recent fits in a conservative manner, we consider here
 \begin{equation}
\frac{\Delta M_{B_s}^{\rm exp}}{\Delta M_{B_s}^{\rm SM}}=1.11\pm0.09\,.
\end{equation}
We obtain the allowed region shown in blue in Fig.~\ref{KKgluon} for real values of the Wilson coefficients and neglecting the bag factor uncertainties related to ${\cal C}_{4,5}^{B_s\bar B_s}$. 


Assuming that the KK gluon has universal flavour-diagonal coupling to the first two generations of quarks,
which is also needed to avoid unacceptably large effects in $K-\bar K$ and/or $D^0-\bar D^0$ mixings~\cite{Calibbi:2020dyg},
our model generates~\footnote{Note that 
our model is only flavour universal with respect to four but not five flavours and does not fulfill the requirements of Sec.~\ref{sec:indepNP}. However, the effect of bottom quarks within the $Q_{4s}$ operator in $L_{K^*\bar{K}^*}$ is $O(\alpha_s)$-suppressed within QCD factorisation and thus the impact of our model on $L_{K^*\bar{K}^*}$
can be mimicked by a shift in ${\cal C}_{4s}$ to a good approximation.} a NP contribution to ${\cal C}_{4s}$ given at the matching scale by
\begin{equation}\label{eq:C4qKK}
 {\cal C}_{4s} =  - \frac{1}{4}\frac{{\Delta _{sb}^L\Delta _{qq}^{L}}}{{\sqrt 2 {G_F}{V_{tb}}V_{ts}^*m_{KK}^2}}\,,
\end{equation}
(and {similarly} for $\tilde{\cal C}_{4{s}}$ with $L$ replaced by $R$). The couplings $\Delta _{sb}^{L,R}$ are defined in Eq.(\ref{lagrangian}) while $\Delta _{qq}^{L,R}$ stand for the corresponding flavour-diagonal couplings to up and down quarks {of the first two generations}.

However, couplings of first generation quarks to KK gluons are strongly constrained by di-jet searches~\cite{Sirunyan:2017ygf}: $(\Delta_{qq}^{L}/m_{KK})^2<(2.2/(10\,{\rm TeV}))^2$. Allowing for NP also in $b\to d$ transitions could increase the effect in $L_{K^*\bar{K}^*}$, but since here the effect is bounded by $B_d-\bar B_d$ mixing, whose constraints are of the same order as $B_s-\bar B_s$ mixing, one can only gain a factor $\approx 2$. Using this maximal coupling for the $\Delta_{qq}^{L}$ couplings and setting the $\Delta_{qq}^R$ couplings to zero, we can see from Fig.~\ref{KKgluon} that a significant amount of fine-tuning is needed to account for $L_{K^*\bar{K}^*}$. 

Alternatively, one could try to explain $L_{K^*\bar{K}^*}$ with a NP contribution in the chirally-flipped coefficient $\tilde{\cal C}_{4{s}}$, given by Eq.~(\ref{eq:C4qKK}) with the  $\Delta _{sb}^L$ and $\Delta _{qq}^{L}$ couplings replaced  by $\Delta _{sb}^R$ and $\Delta _{qq}^{R}$, respectively. In principle, one could exploit the fact that the couplings do not have to respect an $U(2)$ flavour symmetry (since up- and down-type quark couplings are not related via $SU(2)_L$), so that couplings to first-generation quarks could be avoided, which would relax LHC bounds and reduce the fine-tuning needed in $B_s-\bar B_s$ mixing. However, as in the previous case, flavour universality for diagonal couplings to quarks is needed to be able to make use of our expressions for $L_{K^*\bar{K}^*}$. Moreover, according to QCD factorisation, the dominant LO effect in $L_{K^*\bar{K}^*}$ originates from the term in $Q_{4s}$ with down quarks in the bilinear summed over flavours. Therefore, (dominant) right-handed couplings cannot be used to evade LHC bounds and still  fine-tuning in $B_s-\bar B_s$ mixing, like in the case of left-handed couplings, is needed.

As indicated earlier, one could also try to explain $L_{K^*\bar{K}^*}$ with the Wilson coefficient of the chromomagnetic operator $O_{8gs}$. Here an effect of the order of the SM contribution is required. 
${\cal C}_{8gs}$ can only be generated at the loop level and involves necessarily coloured particles for which strong LHC limits exist. Therefore, a value of the order of the SM contribution can only be obtained thanks to chiral enhancement.

A simplified model fulfilling these requirements features two vector-like quarks, one $SU(2)_L$ doublet and one $SU(2)_L$ singlet (with a large coupling $\lambda$ to the SM Higgs doublet) and an additional neutral scalar particle~\cite{Arnan:2019uhr}. In this setup, ${\cal C}_{8gs}$ receives a contribution which scales like $\lambda/(m_b/v) \times v^2/M^2$ w.r.t. the SM, where $M$ is the NP scale. Inevitably an effect in ${\cal C}_{7\gamma s}$ is generated at the matching scale $M$ which however has free sign and magnitude as it depends on the (not necessarily quantized) electric charges of the new fermions and scalar inside the loop. Therefore, the electric charges of the new particles can be chosen in such a way that in ${\cal C}_{7\gamma s}$ (at the $m_b$ scale) the NP contributions to ${\cal C}_{7\gamma s}$ and ${\cal C}_{8gs}$ (taken at the matching scale) cancel. As we need a NP contribution to ${\cal C}_{8gs}$ of the order of the SM one, and ${\cal C}_{7\gamma s}$ at the low scale is known at the $5\%$ level, a tuning of the order of 1/20 is necessary here. 

Both simplified models allow for the possibility of a connection with the $b\to s\ell^+\ell^-$ anomalies. On the one hand, the KK gluon may be part of the particle spectrum of a composite/extra-dimensional model and is then accompanied by a $Z^\prime$ boson. This could explain $b\to s\ell^+\ell^-$ without violating LHC di-lepton bounds~\cite{Aad:2020otl} due to the large $sb$ coupling of the $Z^\prime$ needed to explain $L_{K^*\bar{K}^*}$, leading to NP contributions with the correct sign in both types of anomalies. On the other hand, the model generating a large effect in ${\cal C}_{8g}$ could easily be extended by a vector-like lepton in order to account for $b\to s\ell^+\ell^-$~\cite{Arnan:2019uhr}.

 \section{Conclusions}\label{sec:conclusions}
 
In this article, we have analysed the non-leptonic penguin decays $B_d\to K^{*0}\bar{K}^{*0}$ and $B_s\to K^{*0}\bar{K}^{*0}$, where recent LHCb results indicate striking differences in the longitudinal polarisation of these two modes. This is unexpected since they are related by $U$-spin and should thus have a similar QCD and EW  dynamics (up to tiny corrections due to the down and strange quark masses).
 
 We introduced the $L$-observable as a combination  of polarisation fractions and branching ratios in order to compare the longitudinal amplitudes in both modes, as they can be computed with better theoretical control in a $1/m_b$ expansion such as QCD factorisation.
 We exploited the fact that these penguin-mediated decays exhibit very similar hadronic matrix elements for the ``tree'' and ``penguin'' contributions in the usual decomposition based on CKM factors, so that these contributions are very strongly correlated. This means that the $L$-observable is a measure of $U$-spin breaking between the penguin contributions to $B_d$ and $B_s$ decays, with a deviation from the SM expectation between 2$\sigma$ and $3\sigma$ depending on the specific theoretical framework considered. This observation reinforces and puts on a firmer ground the hint for NP already suspected by  considering the difference between the longitudinal polarisation fractions in these two modes. We performed a detailed error budget analysis for $L_{K^* \bar{K^*}}$ and we found a relatively small impact of infrared divergences coming from weak annihilation and hard-spectator scattering, compared to observables like branching ratios or polarisation fractions involving troublesome transverse amplitudes. 
 
 We then interpreted this deviation in a model-independent approach using the weak effective theory. For simplicity, we allowed NP only in SM Wilson coefficients or their chirally-flipped counterparts. We identified three operators which could accommodate the deviation with NP contributions at most as large as the SM. While ${\cal C}_{1q}$ is already very significantly constrained by other nonleptonic modes and LHCb bounds (up to the point of excluding this solution), the situation is less constrained for the strong penguin coefficient ${\cal C}_{4q}$  and the chromomagnetic one ${\cal C}_{8gq}^{\rm eff}$ where NP contributions of a similar size to the SM one are allowed and could explain the deviation in $L_{K^*\bar{K}^*}$. We discussed examples of simplified NP models that could provide large contributions, at the price of accepting fine tuning to accommodate the bounds on $B_s-\bar B_s$ mixing and $b\to s\gamma$. Interestingly, within a general composite or extra-dimensional model~\cite{Contino:2006nn}, the Kaluza-Klein gluon contribution to the $b\to s$ amplitude in $B_s\to K^{*0}\bar{K}^{*0}$ has the same sign as the $Z^\prime$ contribution to $b\to s \ell^+\ell^-$ w.r.t the SM. Therefore, if one accepts the fine-tuning in $B_s-\bar B_s$ mixing, such models can provide a common explanation of $L_{K^*\bar{K}^*}$ and $b\to s \ell^+\ell^-$ data.

This hint of NP in $L_{K^*\bar{K}^*}$
 could be sharpened with a precise estimate of $U$-spin breaking in the form factors involved,  as they drive the theoretical uncertainty of the SM prediction and their correlation is not known precisely. A comparison of the theoretical and experimental information on the polarisations in $B_s\to K^*\phi$ and $B_d\to K^*\phi$  could also be valuable to check whether a similar tension arises. Complementary information could be obtained also from pseudoscalar-vector and pseudoscalar-pseudoscalar penguin-mediated modes ($K^0\bar{K}^{*0}$ and $K^0\bar{K}^{0}$). Moreover, if the same source of NP is responsible for the suppression of $b\to sq\bar{q}$ versus $b\to dq\bar{q}$ and $b\to s\mu\mu$ versus $b\to  see$, it would be certainly interesting to perform a thorough study of $b\to d\ell^+\ell^-$ modes compared to $b\to s\ell^+\ell^-$ ones, which should be accessible with more data from the LHCb and Belle II experiments. This interplay between non-leptonic and semileptonic rare decays could prove highly beneficial in the coming years to identify new $B$-flavour anomalies and understand their actual origin in terms of physics beyond the SM.

 \section{Acknowledgements}
We thank M. Misiak and E. Lunghi for useful discussions on bounds on the chromomagnetic operator. This work received financial support from the Spanish Ministry of Science, Innovation and Universities (FPA2017-86989-P) and the Research Grant Agency of the Government of Catalonia (SGR 1069) [MA, JM]. IFAE is partially funded by the CERCA program of the Generalitat de Catalunya. JM acknowledges  the financial support by ICREA under the ICREA Academia programme. The work of A.C. is supported by a Professorship Grant (PP00P2\_176884) of the Swiss National Science Foundation. 

\newpage

\appendix

\section{Weak effective theory and QCD factorisation framework}\label{app:WET}

The separation between short and long distances at the scale $\mu_b=m_b$ is performed in the weak effective theory to compute $b$-quark decays within the SM:
\begin{equation}\label{eq:wet}
H_{\rm eff}=\frac{G_F}{\sqrt{2}}\sum_{p=c,u} \lambda_p^{(q)}
 \Big({\cal C}_{1s}^{p} Q_{1s}^p + {\cal C}_{2s}^{p} Q_{2s}^p+\sum_{i=3 \ldots 10} {\cal C}_{is} Q_{is} + {\cal C}_{7\gamma s} Q_{7\gamma s} + {\cal C}_{8gs} Q_{8gs}\Big) \,.
\end{equation}
This effective Hamiltonian describes the quark transitions $b \to u\bar{u}s$, $b \to c\bar{c}s$, $b \to sq'\bar{q}'$ with $q' = u,d,s,c,b$, and $b \to sg$, $b \to s\gamma$.
$Q_{1s,2s}^p$ are the left-handed current-current operators arising from $W$-boson exchange,  $Q_{3s\ldots 6s}$ and
$Q_{7s\ldots 10s}$ are QCD and electroweak penguin operators, and $Q_{7\gamma s}$ and $Q_{8gs}$ are the electromagnetic and chromomagnetic dipole operators.
They are given by~\cite{Beneke:2001ev}:
\begin{align}
Q_{1s}^p &= (\bar p b)_{V-A} (\bar s p)_{V-A} \,,  &Q_{7s} &= (\bar s b)_{V-A} \sum_q\,\frac{3}{2} e_q (\bar q q)_{V+A} \,, \nonumber \\
Q_{2s}^p &= (\bar p_i b_j)_{V-A} (\bar s_j p_i)_{V-A} \,, &Q_{8s} &= (\bar s_i b_j)_{V-A} \sum_q\,\frac{3}{2} e_q (\bar q_j q_i)_{V+A} \,, \nonumber \\
Q_{3s} &= (\bar s b)_{V-A} \sum_q\,(\bar q q)_{V-A} \,, &Q_{9s} &= (\bar s b)_{V-A} \sum_q\,\frac{3}{2} e_q (\bar q q)_{V-A} \,, \nonumber \\
Q_{4s} &= (\bar s_i b_j)_{V-A} \sum_q\,(\bar q_j q_i)_{V-A} \,, &Q_{10s} &= (\bar s_i b_j)_{V-A} \sum_q\,\frac{3}{2} e_q (\bar q_j q_i)_{V-A} \,, \nonumber\\
Q_{5s} &= (\bar s b)_{V-A} \sum_q\,(\bar q q)_{V+A} \,, &Q_{7\gamma s} &= \frac{-e}{8\pi^2}\,m_b\bar s\sigma_{\mu\nu}(1+\gamma_5) F^{\mu\nu} b \,,\nonumber \\
Q_{6s} &= (\bar s_i b_j)_{V-A} \sum_q\,(\bar q_j q_i)_{V+A} \, , &Q_{8gs} &= \frac{-g_s}{8\pi^2}\,m_b\, \bar s\sigma_{\mu\nu}(1+\gamma_5) G^{\mu\nu} b \,,
\label{operators}
\end{align}
where $(\bar q_1 q_2)_{V\pm A}=\bar q_1\gamma_\mu(1\pm\gamma_5)q_2$, 
$i,j$ are colour indices, $e_q$ are the electric charges of the quarks 
in units of $|e|$, and a summation over $q=u,d,s,c,b$ is implied. The NLO Wilson coefficients at the scale $\mu=4.2$ GeV are given in Table~\ref{tab:inputs}.

A similar weak effective theory can be written for the $b\to d$ transition by performing the trivial replacement $s\to d$. Neglecting the difference of mass between the $d$ and $s$ quarks, the SM values of the Wilson coefficients are identical in both cases, and we omit the $d$ or $s$ subscript in Table~\ref{tab:inputs}.

In the SM,
${\cal C}_1^{c}={\cal C}_1^{u}$ is the largest coefficient and it corresponds to the colour-allowed tree-level contribution from the $W$ exchange, whereas ${\cal C}_2^{c}={\cal C}_2^{u}$  is colour suppressed. QCD-penguin operators are numerically suppressed, and the electroweak operators even more so. It proves convenient to define the effective coefficients ${\cal C}_{7\gamma}^{\rm eff}$  and ${\cal C}_{8g}^{\rm eff}$ which are given in the scheme of Ref.~\cite{Beneke:2001ev} as
\begin{eqnarray}
{\cal C}_{7\gamma}^{\rm eff}&=&
{\cal C}_{7\gamma}-\frac{1}{3}{\cal C}_5-{\cal C}_6\,,\\
{\cal C}_{8g}^{\rm eff}&=& {\cal C}_{8g}+{\cal C}_5\,,
\end{eqnarray}
QCD factorisation relies on this weak effective theory
to compute non-leptonic $B$-decay hadronic matrix elements, by 
performing a further separation of scales between $m_b$ and the typical QCD scale, later reinterpreted in terms of a Soft-Collinear Effective Theory (SCET). Following Refs.~\cite{Beneke:2003zv,Beneke:2006hg} 
 and using the same notation as in this reference, we have for the vector modes for a given polarisation:
\begin{eqnarray}\label{main}
&&\!\!\!T(\bar{B}_d\to \bar{K}^{*0}K^{*0})=A_{\bar{K}^*K^*}
   [\alpha_4^u-\frac{1}{2}\alpha_{4,EW}^u+\beta_3^u+\beta_4^u-\frac{1}{2}\beta_{3,EW}^u-\frac{1}{2}\beta_{4,EW}^u] \nonumber \\
  && \qquad \qquad \qquad \qquad \, \,  +A_{K^*\bar{K}^*}[\beta_4^u-\frac{1}{2}\beta_{4,EW}^u]\,, \nonumber \\
&&\!\!\! P(\bar{B}_d\to \bar{K}^{*0}K^{*0})=A_{\bar{K}^*K^*}
   [\alpha_4^c-\frac{1}{2}\alpha_{4,EW}^c+\beta_3^c+\beta_4^c-\frac{1}{2}\beta_{3,EW}^c-\frac{1}{2}\beta_{4,EW}^c] \nonumber \\
   && \qquad \qquad \qquad \qquad \, \, +A_{K^*\bar{K}^*}[\beta_4^c-\frac{1}{2}\beta_{4,EW}^c]\,,\nonumber \\
&&\!\!\!T(\bar{B}_s\to \bar{K}^{*0}K^{*0})=A_{\bar{K}^*K^*}[\beta_4^u-\frac{1}{2}\beta_{4,EW}^u]\nonumber\\
&& \qquad \qquad \qquad \qquad \, \,  +A_{K^*\bar{K}^*}
   [\alpha_4^u-\frac{1}{2}\alpha_{4,EW}^u+\beta_3^u+\beta_4^u-\frac{1}{2}\beta_{3,EW}^u-\frac{1}{2}\beta_{4,EW}^u] \,,\nonumber\\
&&\!\!\! P(\bar{B}_s\to \bar{K}^{*0}K^{*0})=A_{\bar{K}^*K^*}[\beta_4^c-\frac{1}{2}\beta_{4,EW}^c]\nonumber\\
&& \qquad \qquad \qquad \qquad \, \,  +A_{K^*\bar{K}^*}[\alpha_4^c-\frac{1}{2}\alpha_{4,EW}^c+\beta_3^c+\beta_4^c-\frac{1}{2}\beta_{3,EW}^c-\frac{1}{2}\beta_{4,EW}^c] \,. \nonumber \\
 \end{eqnarray}
 The coefficients $\alpha$ and $\beta$ involve form factors and convolutions of perturbative kernels with light-cone distribution  amplitudes multiplied by the Wilson coefficients of the weak effective Hamiltonian.
The difference between $\alpha_i^u$ and $\alpha_i^c$ occurs from the ${\cal O}(\alpha_s)$ penguin contractions in $P_4^p$ and $P_6^p$, and specifically from the loops with $u$ or $c$ quarks and a $W$ exchange (so that these contributions come with factors $\alpha_s/(4\pi)$ and ${\cal C}_1^{c}$). This comes from the fact that the effective Hamiltonian has a specific structure in the SM: only two types of four-fermion operators $O_1^p$ and $O_2^p$ ($p=u,c$) involve explicitly different $\lambda_p^{(q)}$, whereas the other operators treat all quarks on the same footing, they come from top loops and are accompanied with a CKM term $\lambda_t^{(q)}=-\lambda_u^{(q)}-\lambda_c^{(q)}$ leading to an identical contribution to $T$ and $P$.

As discussed in Refs.~\cite{DescotesGenon:2006wc,DescotesGenon:2007qd,DescotesGenon:2011pb}, this explains why the quantity $\Delta$ defined in Eq.~(\ref{delta})
can be computed safely within QCD factorisation for penguin mediated decays because of the cancellation of long-distance contributions. As a consequence of this cancellation, only penguin contractions contribute to $\Delta$,  as can be seen by inspection of the formulae above, leading to the following very simple expression within QCD factorisation:
\begin{equation}
    \Delta = A^{Q}_{M_1M_2}\frac{C_F\alpha_s}{4\pi N}{\cal C}_{1}[\bar{G}_{M_2}(m^2_c/m^2_b)-\bar{G}_{M_2}(0)]\,,
\end{equation}
where the normalisation $A^{Q}_{M_1M_2}$ is defined as:
\begin{equation}
    A^{Q}_{M_1M_2}=\frac{G_F}{\sqrt{2}}m^2_{B_q}f_{M_2}A^{B_q\to M_1}(0)\,,
\end{equation}
and $\bar{G}_{M_2}$ is the penguin function defined in Ref.~\cite{DescotesGenon:2006wc}.

\begin{table}[h]
	\begin{center}
\tabcolsep=1.26cm\begin{tabular}{|c|c|c|}
\hline\multicolumn{3}{|c|}{$B_{d,s}$ Distribution Amplitudes (at $\mu=1$ GeV)~\cite{Khodjamirian:2020hob,Ball:2006nr} } \\ \hline
$\lambda_{B_d} $ [GeV]&$\lambda_{B_s}/\lambda_{B_d}$& $\sigma_B$\\
\hline
$0.383\pm0.153$&$1.19\pm0.14$&$1.4\pm0.4$\\ \hline
\end{tabular}

\vskip 1pt

\tabcolsep=0.712cm\begin{tabular}{|c|c|c|c|}
\hline\multicolumn{4}{|c|}{$K^*$ Distribution Amplitudes (at $\mu=2$ GeV)~\cite{Ball:2007rt}}  \\ \hline
$\alpha_1^{K^*}$&
$\alpha_{1,\perp}^{K^*}$&
$\alpha_2^{K^*}$&
$\alpha_{2,\perp}^{K^*}$\\
\hline
$0.02\pm0.02$&$0.03\pm0.03$&$0.08\pm0.06$&$0.08\pm0.06$\\ \hline
\end{tabular}

\vskip 1pt

\tabcolsep=0.496cm\begin{tabular}{|c|c|c|c|}
\hline\multicolumn{4}{|c|}{Decay Constants (at $\mu=2$ GeV)~\cite{Aoki:2019cca,Straub:2015ica,Allton:2008pn}}  \\ \hline
$f_{B_d}$&$f_{B_s}/f_{B_d}$&$f_{K^*}$&$f^\perp_{K^*}/f_{K^*}$\\
\hline
$0.190\pm0.0013$&$1.209\pm0.005$&$0.204\pm0.007$&$0.712\pm0.012$\\ \hline
\end{tabular}

\vskip 1pt

\tabcolsep=0.52cm\begin{tabular}{|c|c|c|c|}
\hline\multicolumn{4}{|c|}{$B_{d,s}\to K^*$ form factors~\cite{Straub:2015ica} and B-meson lifetimes (ps)}  \\ \hline
$A_0^{B_s}(q^2=0)$&$A_0^{B_d}(q^2=0)$ & $\tau_{B_d}$ & $\tau_{B_s}$\\
\hline
$0.314 \pm 0.048$ &$ 0.356 \pm 0.046$ & $1.519\pm0.004$ & $1.515\pm0.004$ \\ \hline
\end{tabular}

\vskip 1pt

\tabcolsep=0.515cm\begin{tabular}{|c|c|c|c|c|}
\hline
\multicolumn{4}{|c|}{Wolfenstein parameters~\cite{Charles:2004jd} }  \\ \hline
$A$ & $\lambda$ & $\bar\rho$ &
$\bar\eta$\\ \hline
$0.8235^{+0.0056}_{-0.0145}$ &$	0.22484^{+0.00025}_{-0.00006}$&$0.1569^{+0.0102}_{-0.0061}$&$0.3499^{+0.0079}_{-0.0065}$\\
\hline
\end{tabular}

\vskip 1pt

\tabcolsep=0.496cm\begin{tabular}{|c|c|c|c|c|c|}
	\hline
	\multicolumn{6}{|c|}{QCD scale and masses [GeV]}\\
		\hline 
 $\bar{m}_b(\bar{m}_b)$  & $m_b/m_c$ & $m_{B_d} $& $m_{B_s} $& $m_{K^*} $ & $\Lambda_{{\rm QCD}}$
      \\ \hline
      $4.2$  & $4.577\pm0.008$  & $5.280$ & $5.367$&$0.892$&$0.225$
      \\ \hline
\end{tabular}
\vskip 1pt
\begin{tabular}{|c|c|c|c|c|c|}
\hline
	\multicolumn{6}{|c|}{SM Wilson Coefficients (at $\mu=4.2$ GeV)}\\
		\hline 
${\cal C}_1$ &  ${\cal C}_2$ & ${\cal C}_3$ & ${\cal C}_4$ &  ${\cal C}_5$ & ${\cal C}_6$
      \\ \hline
 1.082 & -0.191 & 0.013 & -0.036 & 0.009 &  -0.042
      \\ \hline
 ${\cal C}_{7}/\alpha_{em}$ & ${\cal C}_{8}/\alpha_{em}$ & ${\cal C}_{9}/\alpha_{em}$ &  ${\cal C}_{10}/\alpha_{em}$ & ${\cal C}^{\rm eff}_{7\gamma}$ &  ${\cal C}^{\rm eff}_{8g}$
      \\ \hline
 -0.011 & 0.058 & -1.254 & 0.223 & -0.318 & -0.151 
      \\ \hline
		\end{tabular}
		\caption{Input parameters used to determine the SM predictions.}
		\label{tab:inputs}
	\end{center}
\end{table}

\vspace{-0.65cm}

\section{Semi-analytical expressions}
\label{app:semianalytical}

In the following we provide the key elements to construct a semi-analytical expression of $L_{K^*\bar{K}^*}$. Specifically we give $P_s$ and $P_d$ in terms of Wilson coefficients and the parameters $X_H$ and $X_A$. $\kappa$ is given in Eq.~(\ref{eq:kappa}) and the last bracket in Eq.~(\ref{eq:LKstKstDeltaP}) has a negligible impact and can be taken to be conservative $0.99 \pm 0.01$. We have followed the corrected expression of Ref.~\cite{Bartsch:2008ps} for the modelling of the weak annihilation in terms of $X_A$.

\vspace{-0.5cm}
\begin{align}
\nonumber
  10^7\times P_d&=i0.076 {\cal C}_{7\gamma }^{\rm eff} - i8.8 {\cal C}^{\rm eff}_{8g} + ((2.6 - i1.8 ) + i0.13  X_A  - i0.041  X_A^2 - i0.025 X_H) {\cal C}_1^{c}\\ \nonumber
     &+ ((-0.045 + i0.39) - i 0.61 X_A +i0.16 X_A^2 + i 0.035 X_H ){\cal C}_2^c \\ \nonumber
     &+ ((15.5 + i38.9 ) + i 0.31 X_A + i 0.25 X_A^2 +i 3.8  X_H) {\cal C}_3 \\ \nonumber
     &+ ((11.0 + i156.9 ) + i 0.25 X_A +i 0.96 X_A^2 - i 0.54 X_H) {\cal C}_4 \\ \nonumber
     &+ ((-7.4 - i 7.2) + i 9.2 X_A - i 3.3 X_A^2 + i 0.11 X_H) {\cal C}_5 \\ \nonumber
     &+ ((11.0 - i19.9) + i 27.7 X_A - 8.9 X_A^2 + i 0.24 X_H) {\cal C}_6 \\ \nonumber
     &+ ((3.7 + i3.8 ) - i 4.7 X_A + i 1.7 X_A^2 + i 0.00042 X_H) {\cal C}_7 \\ \nonumber
     &+ ((  i6.9) - i 15.7 X_A +i 5.0 X_A^2 - i 0.008 X_H) {\cal C}_8 \\ \nonumber
     &+ ((-6.4 - i19.4) - i 0.55 X_A - i 0.041 X_A^2 - i 1.9 X_H) {\cal C}_9 \\ 
     &+ (- i 81.9 - 1.4 X_A - i 0.15 X_A^2 + i 0.32 X_H) {\cal C}_{10}\,, \\ \nonumber
     & \\ \nonumber
   10^7\times P_s&=i0.069 {\cal C}_{7\gamma}^{\rm eff} - i8.0 {\cal C}^{\rm eff}_{8g} + ((2.4 - i1.7 ) + i0.16  X_A - i0.049  X_A^2 - i0.026 X_H) {\cal C}_1^{c}\\ \nonumber
     &+ ((-0.041 + i0.45) - i 0.74 X_A +i0.1 X_A^2 + i 0.037 X_H ){\cal C}_2^c \\ \nonumber
     &+ ((14.2 + i36.4 ) + i 0.37 X_A + i 0.3 X_A^2 +i 3.9  X_H) {\cal C}_3 \\ \nonumber
     &+ ((10.0 + i142.7 ) + i 0.31 X_A +i 1.2 X_A^2 - i 0.56 X_H) {\cal C}_4\\  \nonumber
     &+ ((-6.7 - i 7.7) + i 11.1 X_A - i 3.9 X_A^2 + i 0.11 X_H) {\cal C}_5 \\ \nonumber
     &+ ((10.0 - i21.7) + i 33.5 X_A - 10.8 X_A^2 + i 0.25 X_H) {\cal C}_6 \\ \nonumber
     &+ ((3.4 + i4.0 ) - i 5.7 X_A + i 2.0 X_A^2 + i 0.00043 X_H) {\cal C}_7 \\ \nonumber
     &+ ((i8.3) - i 19.0 X_A +i 6.0 X_A^2 - i 0.008 X_H) {\cal C}_8 \\ \nonumber
     &+ ((-5.8 - i18.1) - i 0.66 X_A - i 0.049 X_A^2 - i 2.0 X_H) {\cal C}_9\\
     &+ (- i 74.3 - 1.7 X_A - i 0.18 X_A^2 + i 0.33 X_H) {\cal C}_{10}\,.
\end{align}

\section{Sensitivity to New Physics}\label{app:plots}

\begin{figure}[h!]
\centering
	\includegraphics[width=0.64\textwidth]{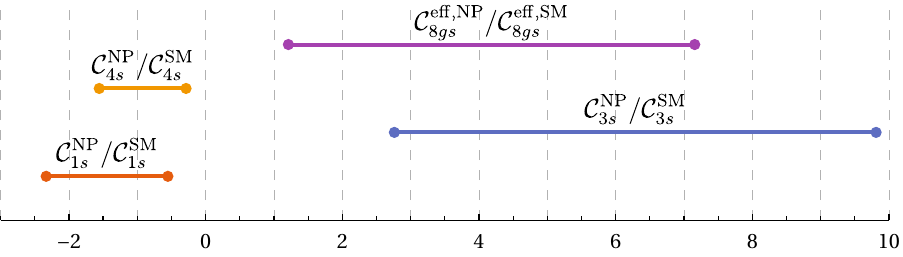} 
\caption{\label{fig:1sigmaintervalsforCiNP} 
 $1\sigma$ intervals for the NP contribution to Wilson coefficients needed to explain $L_{K^{*}\bar{K}^{*}}$, normalised to their SM value.}
\end{figure}

We show how NP contributions can help to reduce the tension between theory and experiment for $L_{K^{*}\bar{K}^{*}}$, completing the results shown in Fig.~\ref{fig:C1sC4sC8gs} discussed in Sec.~\ref{sec:indepNP}. In Fig.~\ref{fig:1sigmaintervalsforCiNP} 
we show the $1\sigma$-range for the NP contribution to each Wilson coefficient that is able to explain the experimental value of $L_{K^*\bar{K}^*}$, normalised to its SM value.
\vspace{0.25cm}

\begin{figure*}
\centering
\includegraphics[width=0.40\textwidth]{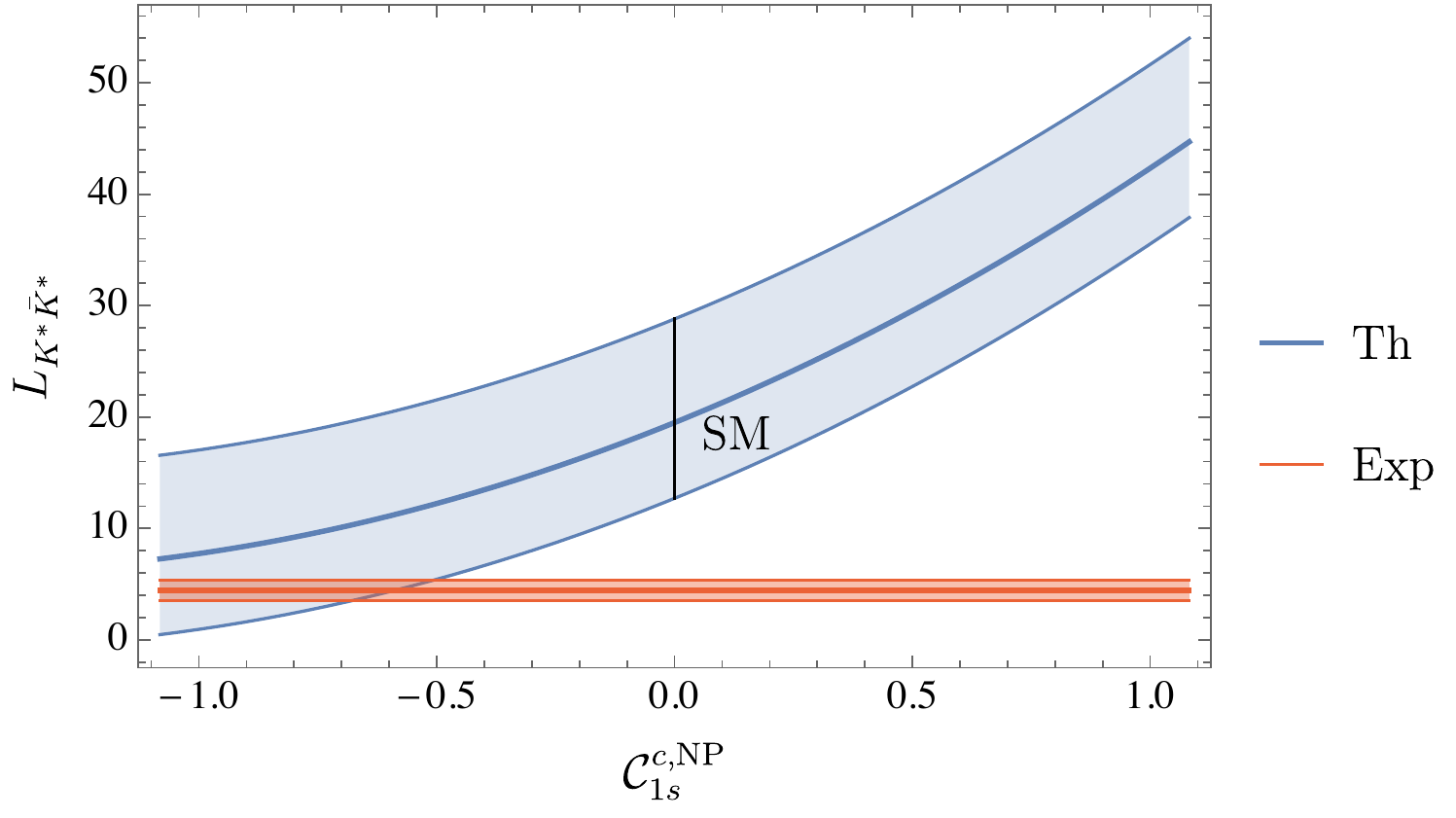}\qquad
\includegraphics[width=0.40\textwidth]{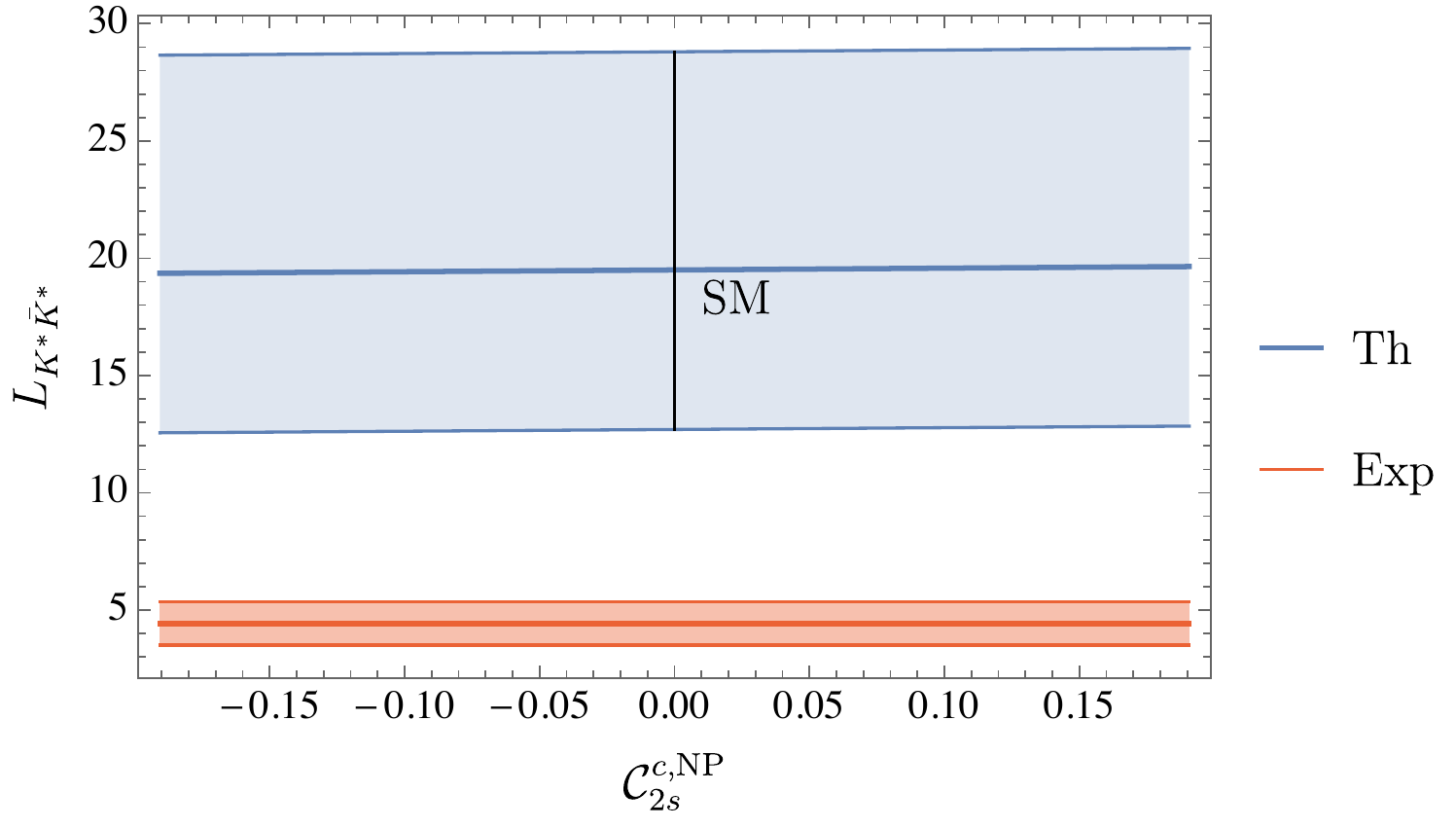}\qquad
\includegraphics[width=0.40\textwidth]{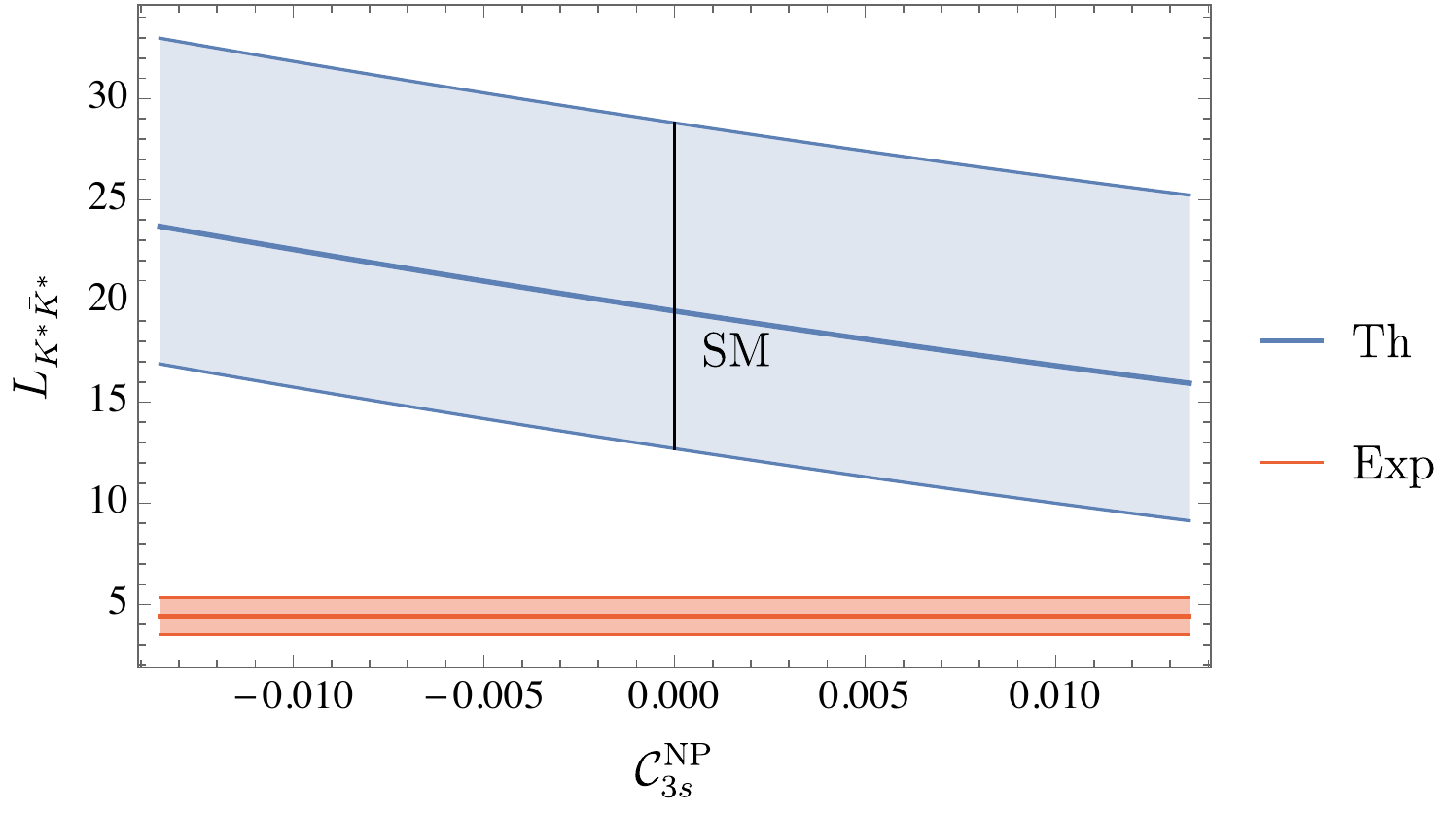}\qquad
\includegraphics[width=0.40\textwidth]{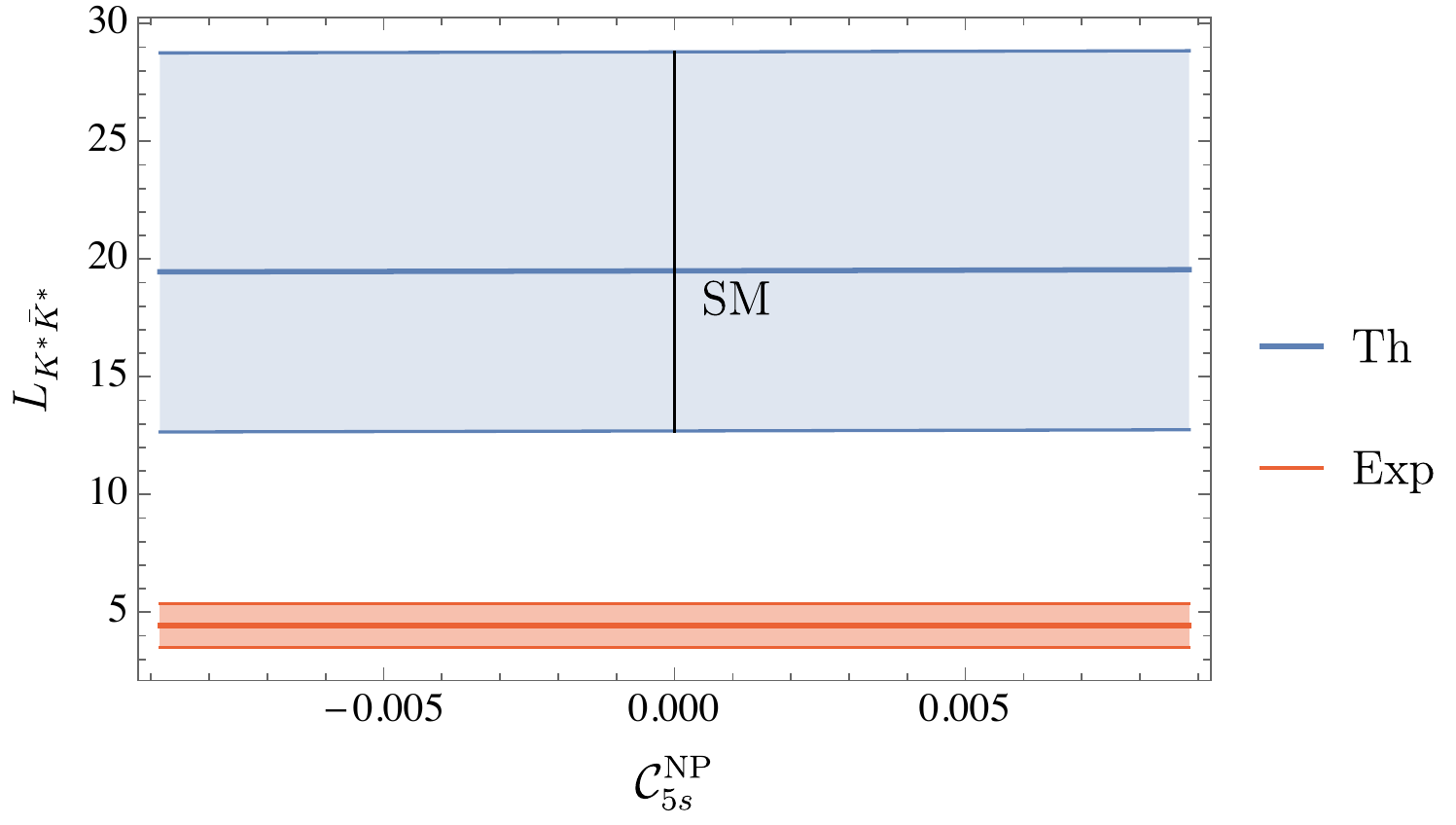}\qquad
\includegraphics[width=0.40\textwidth]{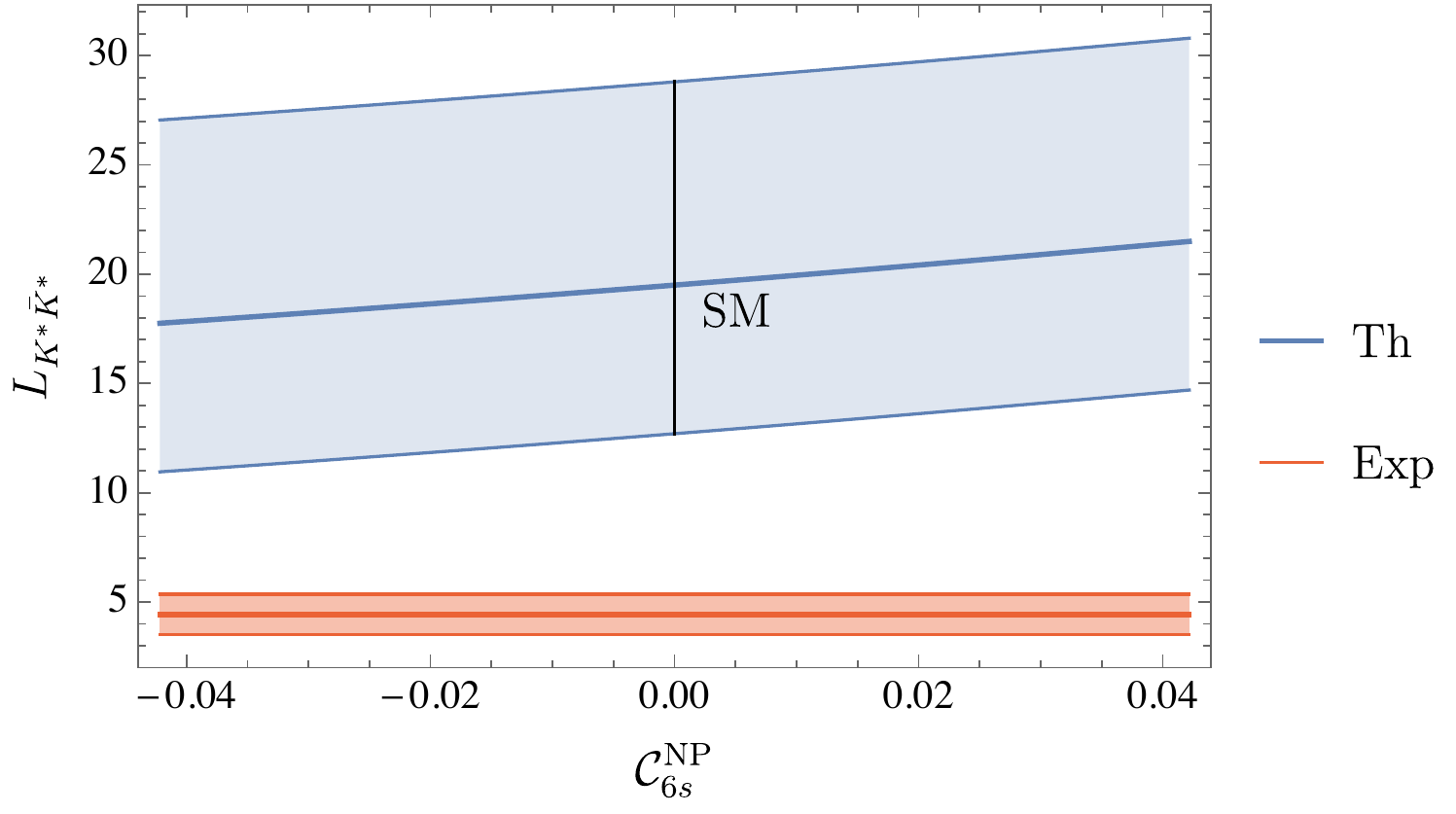}\qquad
\includegraphics[width=0.40\textwidth]{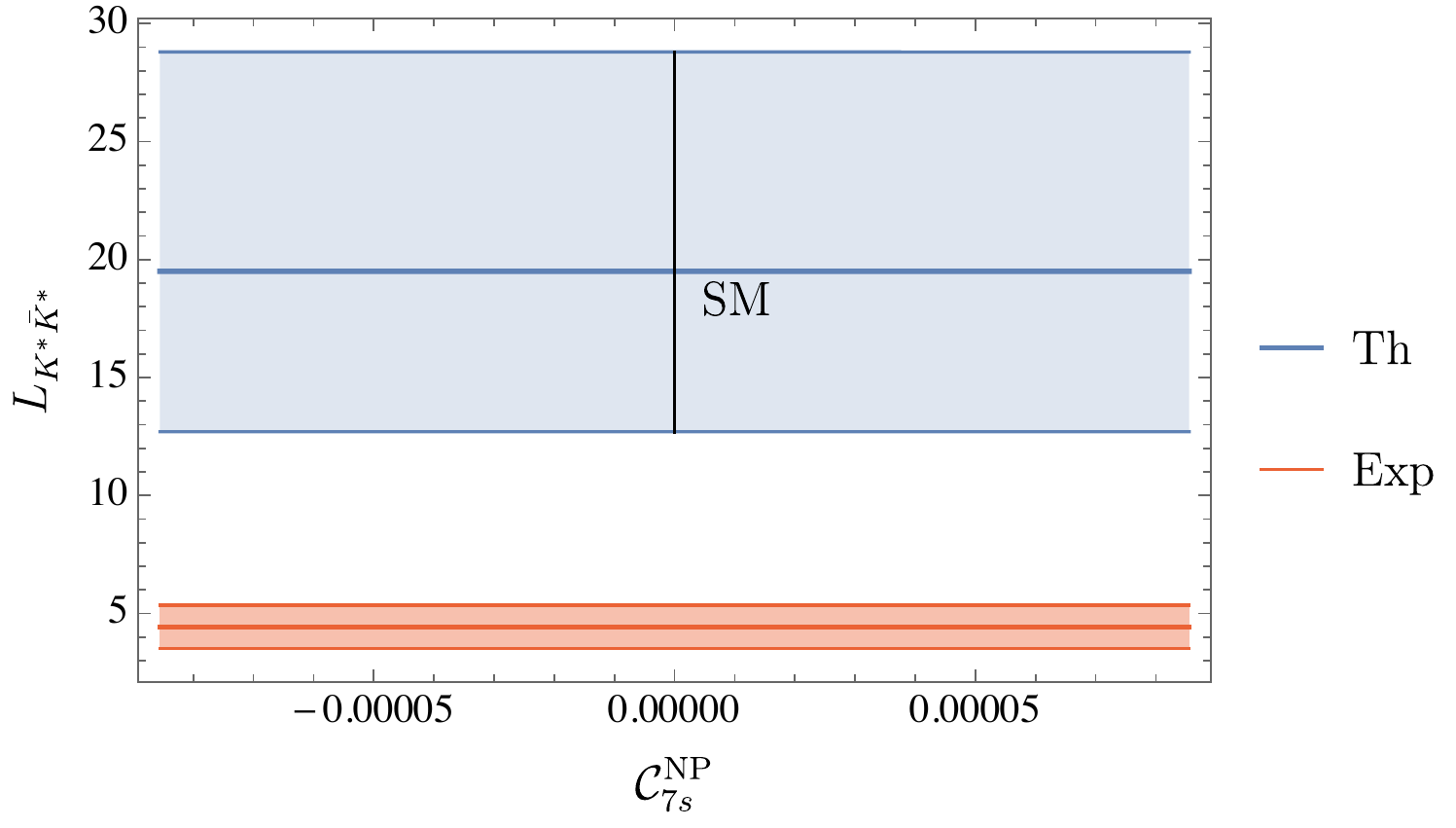}\qquad
\includegraphics[width=0.40\textwidth]{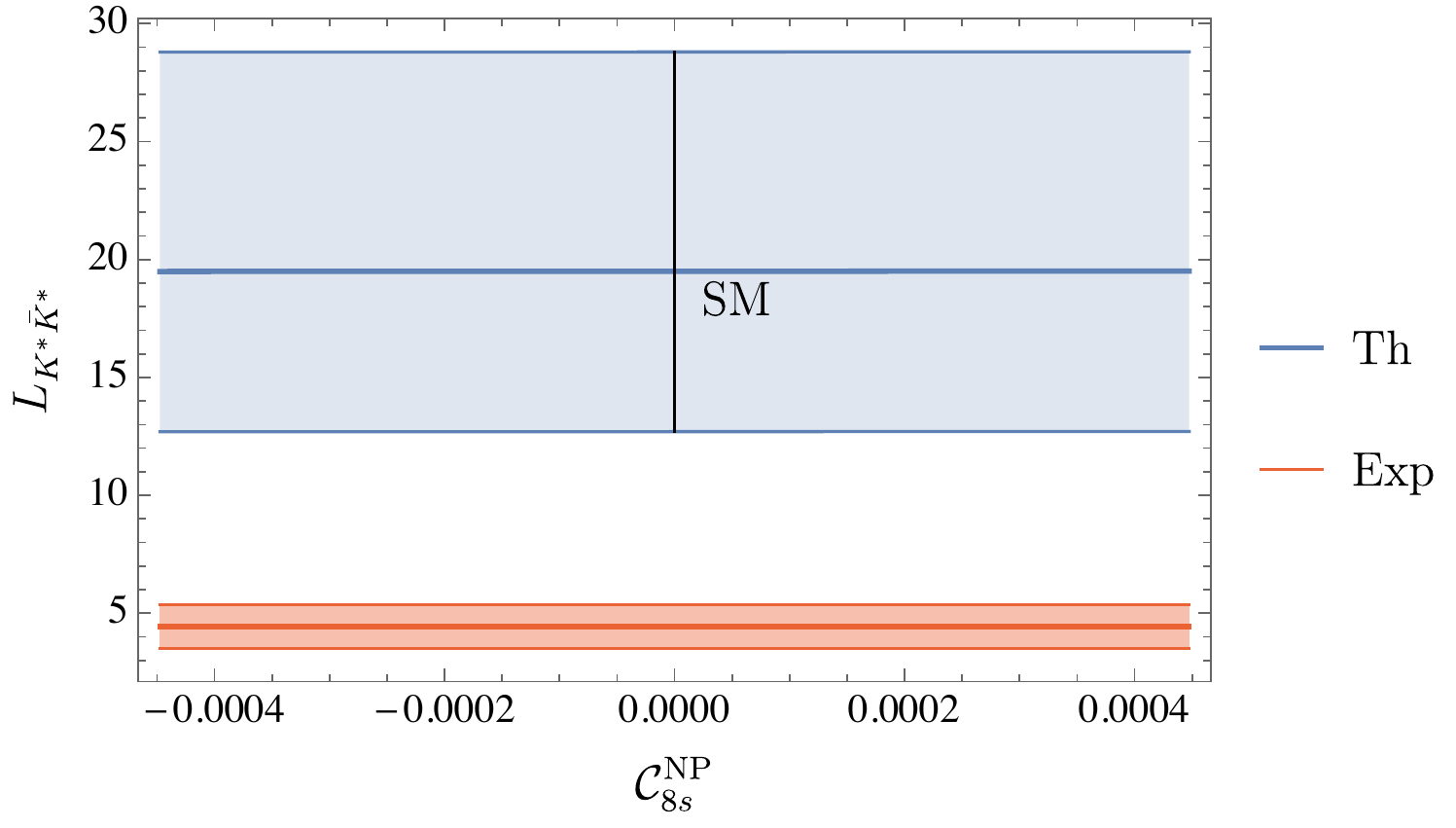}\qquad
\includegraphics[width=0.40\textwidth]{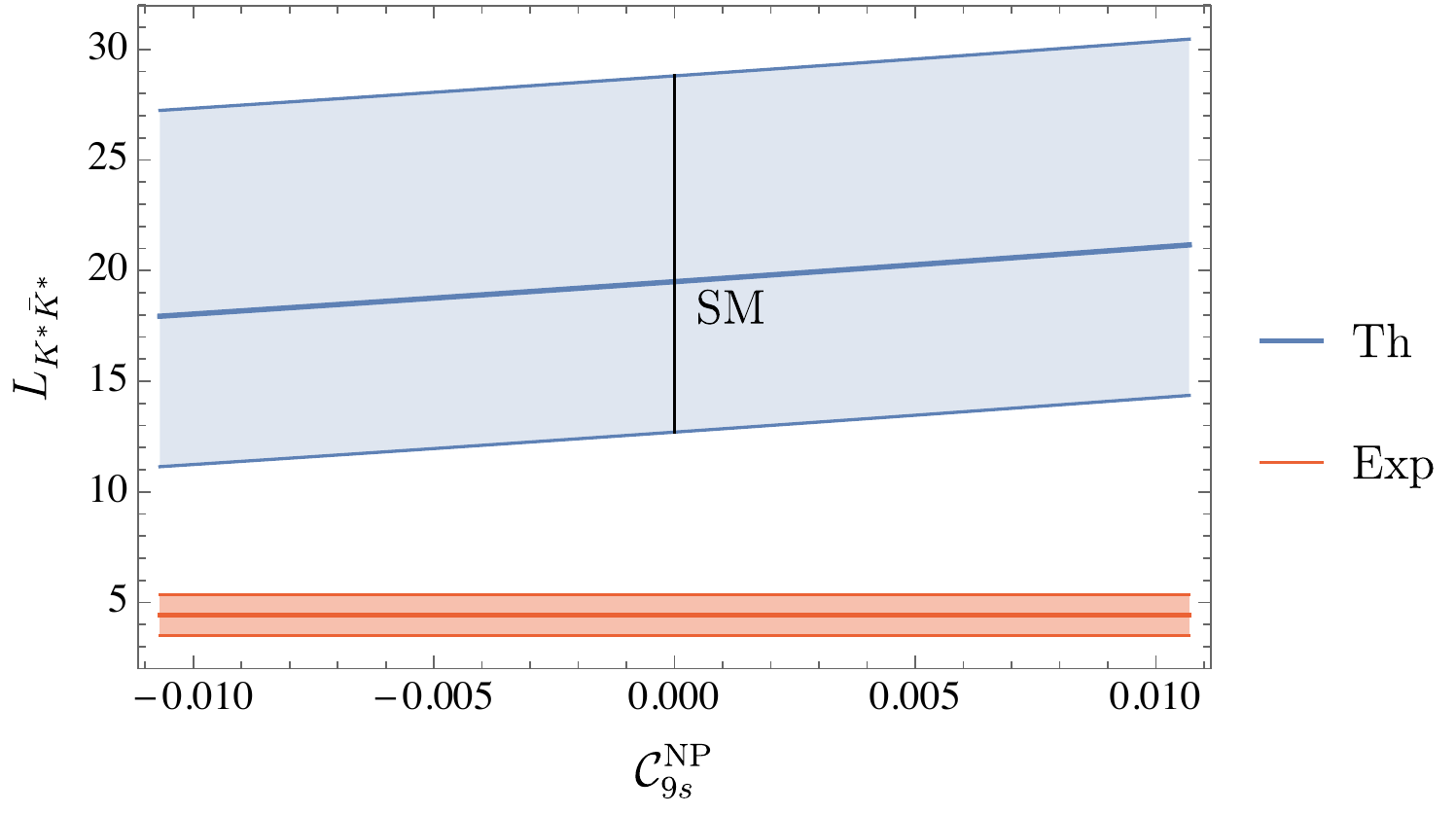}\qquad
\includegraphics[width=0.40\textwidth]{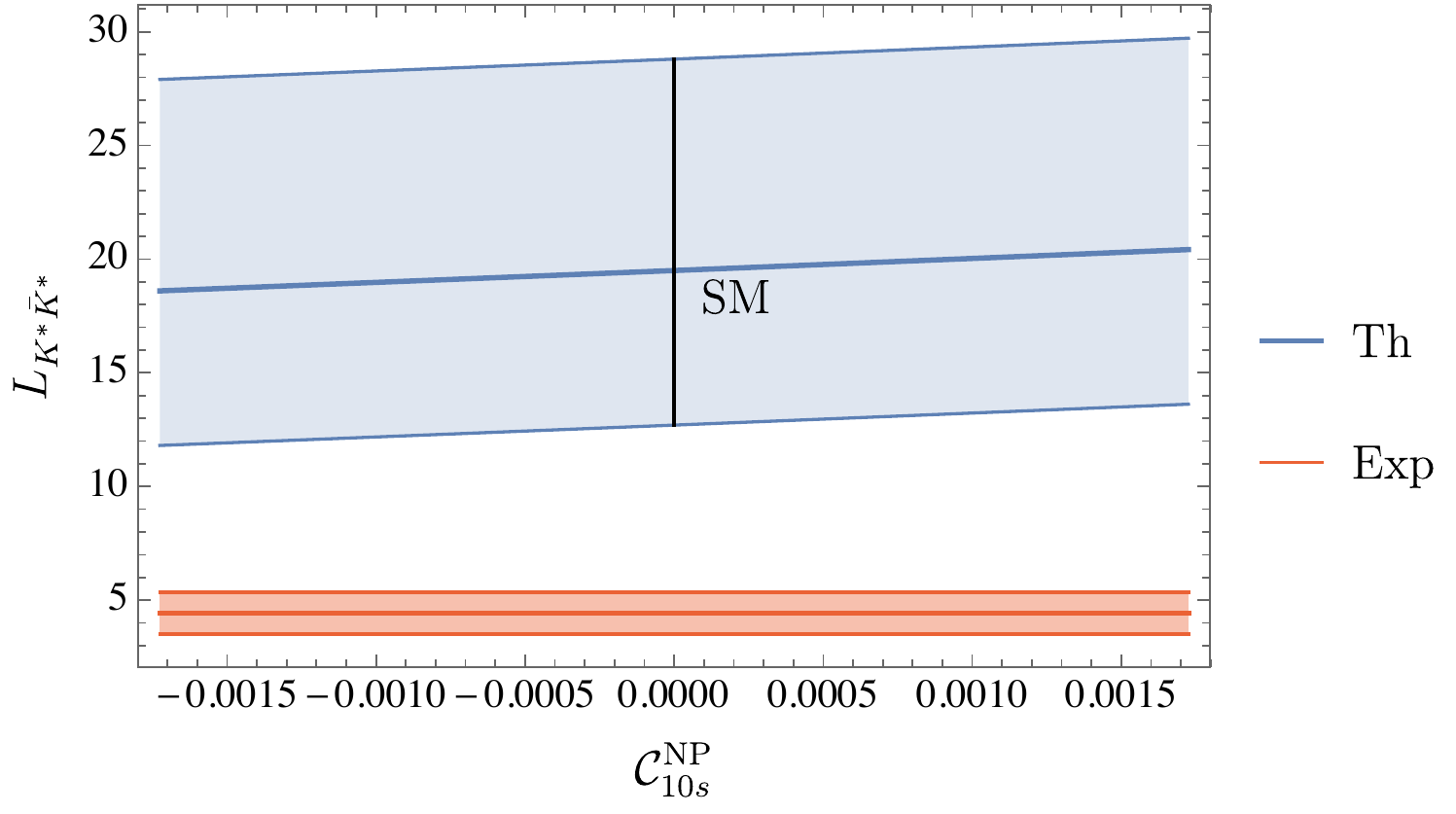}\qquad
\includegraphics[width=0.40\textwidth]{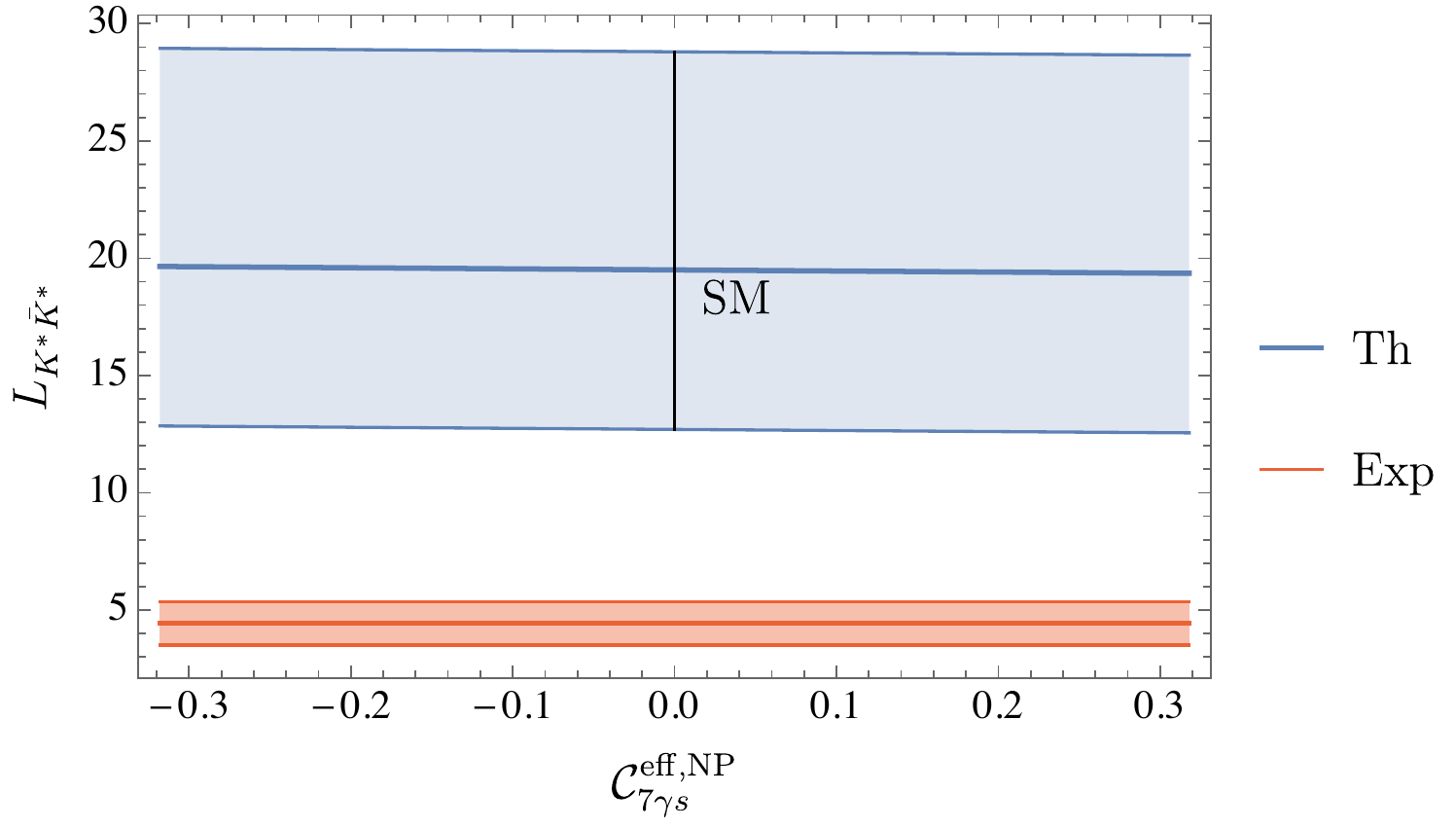}
\caption{\label{fig:s}Sensitivity of $L_{K^{*}\bar{K}^{*}}$ to individual contributions of NP in all different ${\cal C}^{\rm NP}_{is}$. For each coefficient, the range of variation considered for the NP contribution corresponds to 100\% of its SM value.}
\end{figure*}



\clearpage
\bibliography{main}

\end{document}